\documentclass[12pt]{article}
\usepackage{amsmath}
\usepackage{amsfonts}

\advance \textheight by \topskip
\advance \textheight by 1pt
\makeatother
\def\bea{\begin{eqnarray}}
\def\ena{\end{eqnarray}}
\def\non{\nonumber}

\newtheorem{prop}{Proposition}
\newtheorem{theorem}{Theorem}
\newtheorem{defn}{Definition}
\newtheorem{lemma}{Lemma}
\newtheorem{cor}{Corollary}

\def\wt{{\rm wt}\,}

\def\bxi{\mbox{\boldmath $\xi$}}

\newcommand{\bc}[2]{
\left(
\begin{array}{c}{#1}\\{#2}\end{array}
\right)}

\title{
Tau Function Approach to Theta Functions
}

\author{
Atsushi Nakayashiki\thanks{
e-mail: atsushi@tsuda.ac.jp}\\
Department of Mathematics,\\
Tsuda College\\
\\
\\
 \\
}

\date{}
\begin{document}

\maketitle
\thispagestyle{empty}
\vskip15mm

\begin{abstract}
We study theta functions of a Riemann surface of genus $g$  from the view point of $\tau$-function of a hierarchy of soliton equations. We study two kinds of series expansions.  One is the Taylor expansion at any point of the theta divisor.
We describe the initial term  of the expansion  by the Schur function corresponding to the partition determined by the gap sequence of a certain flat line bundle.  The other is the expansion 
of the theta function and its certain derivatives in one of the variables on the Abel-Jacobi images of 
 $k$ points on a Riemann surface with $k\leq g$. We determine the initial term of the expansion as certain derivatives of the theta function successively.
 As byproducts, firstly we obtain a refinement of  Riemann's singularity theorem. Secondly 
we determine normalization constants of higher genus sigma functions of a Riemann surface, 
defined by Korotkin and Shramchenko, 
 such that they become modular invariant.
\end{abstract}
\clearpage
\pagestyle{plain}

\setcounter{page}{1}

\section{Introduction}
The notion of $\tau$-function of an integrable hierarchy of soliton equations is considered as 
an extension of the notion of theta function. This is not only because the $\tau$-function 
becomes a theta function in a special case but also because the structural similarities like 
addition formulae \cite{SS} which are equivalent to the the hierarchy itself \cite{TT,Shig}. 
More concretely let us consider the KP-hierarchy which is considered to be a universal system of
 integrable differential equations in the sense that various soliton equations are derived as special cases of it. 
Its solution space is determined by Sato \cite{SS,SN}
as a certain infinite dimensional Grassmann manifold called the universal Grassmann manifold (UGM). 
In other words there is a one to one correspondence between solutions of the KP-hierarchy($\tau$-functions) and points of UGM.

Sato's theory on the KP-hierarchy was successfully applied to  the study of Novikov's conjecture \cite{No, Mul,Shi}. The results show that theta functions of algebraic curves  are characterized by the KP-hierarchy among those of principally polarized Abelian varieties (ppAv). It means that tau functions of the KP-hierarchy 
supplied by the general properties of theta functions of ppAv can produce all the properties of theta functions 
corresponding to Riemann surfaces. Therefore it is quite natural to study theta functions by way of $\tau$-functions. 

However it seems that a $\tau$-function approach in the study of theta functions is not fully 
developed yet.  In papers \cite{N2,NY,EEG,AN} we have studied the higher genus sigma functions 
of various algebraic curves from the view point of $\tau$-functions.
The aim of this paper is to develop these researches further and to add examples which are effectively studied by an approach of $\tau$-functions.

We study two kinds of series expansions of the theta function of a Riemann surface. One is the series expansion at any point on the theta divisor. 
We determine the initial term, with respect to certain weight,  of the expansion as the Schur function 
with the partition determined from the gap sequence of the flat line bundle corresponding to a point 
on the theta divisor. The other is the expansion of the theta function and certain derivatives of the theta 
function with respect to one of the variables 
on the Abel-Jacobi images of $k$ points on a Riemann surface of genus $g$ with $k\leq g$. We determine 
the initial term of the expansion as a certain explicit derivative of the theta function. 

As a consequence of the study on the expansions we get an extension and a refinement of Riemann's singularity theorem.
As another corollary we determine normalization constants of higher genus sigma 
functions of  a Riemann surface introduced by Korotkin and Shramchenko \cite{KS} so that they are modular 
invariant. In their paper the modular invariance of sigma functions is proved up to multiplications 
of certain roots of unity. We propose apparently different normalization constants from theirs and 
prove the modular invariance. Let us explain our results in more detail.

Let $X$ be a compact Riemann surface of genus $g$, $\{\alpha_i,\beta_i\}$ a canonical 
homology basis, $\Omega$ the normalized period matrix and  $\theta(Z|\Omega)$ Riemann's 
theta function. We consider the data $(X, \{\alpha_i,\beta_i\}, p_\infty,e)$ consisting of 
$X$, $\{\alpha_i,\beta_i\}$ as above, a point $p_\infty$ on $X$ and a point $e$ of the theta 
divisor. We associate a partition to such a data as follows.

To this end we need a notion of gaps of a line bundle.
Let  $L$ be a holomorphic line bundle on $X$ of degree zero, which we call a flat line bundle. A non-negative integer $n$ is called a gap of $L$ at $p_\infty$ if there does not exist a meromorphic section of $L$ which is holomorphic 
on $X\backslash\{p_\infty\}$ and has a pole of order $n$ at $p_\infty$. By Riemann-Roch it can be 
easily proved that there are exactly $g$ gaps in $\{0,1,...,2g-1\}$ for any $(L,p_\infty)$.
Let $\delta$ be Riemann's constant,  $L_{e+\delta}$ the flat line bundle corresponding to the point $e+\delta$ on the Jacobian and 
\bea
&&
b_1<\cdots<b_g,
\non
\\
&&
w_1<\cdots<w_g,
\non
\ena
 the gaps of $L_{e+\delta}$ and $L_0$ at $p_\infty$ respectively, where $L_0$ is the trivial line bundle. 
We define the partition $\lambda=(\lambda_1,...,\lambda_g)$ by
\bea
&&
\lambda=(b_g,b_{g-1},...,b_1)-(g-1,g-2,...,0),
\non
\ena
and consider  the Schur function $s_\lambda(t)$, $t=(t_1,t_2,...)$. The special property of $s_\lambda(t)$ 
 is that it depends only on $t_{w_i}$, $1\leq i\leq g$.  This property is crucial when we connect it to the theta function.

To give a relation of $s_\lambda(t)$ with the theta function  
we need to make a change of variables which is 
given by a certain non-normalized period matrix. To define it 
we specify a local coordinate $z$ around $p_\infty$. Then there is a basis $du_{w_i}$, $1\leq i\leq g$, of holomorphic one forms 
which has the expansion at $p_\infty$ of the form
\bea
&&
du_{w_i}=(z^{w_i-1}+O(z^{w_i}))dz,
\qquad
1\leq i\leq g.
\non
\ena
It is not unique. We take any one of them. Let $2\omega_1$ be the $\{\alpha_i\}$ period matrix 
of $\{du_{w_i}\}$.  Let $u={}^t(u_{w_1},...,u_{w_g})$. 
We assign the weight $i$ to variables $u_i$ and $t_i$. Then  the Schur function $s_{\lambda}(t)$ becomes a 
weight-homogeneous polynomial with the weight $|\lambda|=\lambda_1+\cdots+\lambda_l$ for 
$\lambda=(\lambda_1,...,\lambda_l)$.

We prove that the theta function has the expansion of the form 
\bea
&&
C\theta\bigl((2\omega_1)^{-1}u+e|\Omega\bigr)=s_\lambda(t)|_{t_{w_i}=u_{w_i}}+\text{higher weight terms},
\label{intro-1}
\ena
for some constant $C$ which is given explicitly by a theta constant (see Theorem \ref{expansion-theta}).

Next we study the expansion of  the function 
$\theta(p_1+\cdots+p_g+e|\Omega)$ in $z_g=z(p_g)$ 
where $p_i$ inside the theta function denotes the image of $p_i$  by the Abel-Jacobi map 
with the base point $p_\infty$.
To describe the results we need some sequence of numbers which we call a-sequence.

Let $0\leq b_1^\ast<b_2^\ast<\cdots$ be non-gaps of $L_{e+\delta}$ at $p_\infty$.
For $0\leq k\leq g-1$ define $m_k$ by 
\bea
&&
m_k=\sharp\{i\,|\, b_i^\ast <g-k\}
\non
\ena
and $a^{(k)}_i$, $1\leq i\leq m_k$, by 
\bea
&&
(a^{(k)}_1,...,a^{(k)}_{m_k})=(b_{g-k},b_{g-k-1},...,b_{g-k-m_k+1})-(b_1^\ast,...,b_{m_k}^\ast).
\non
\ena
Any $a^{(k)}_i$ is proved in $\{w_j\}$. In general for a non-empty subset $I=\{i_1,...,i_l\}$ 
we set 
\bea
&&
\partial_I=\partial_{u_{i_1}}\cdots \partial_{u_{i_l}}, \qquad \partial_{u_i}=\frac{\partial}{\partial u_i}.
\non
\ena
and set $\partial_I=1$ for $I=\phi$. 
Let $A_k=\{a^{(k)}_i\}$ for $k\geq 1$ and $A_0=\phi$. 
We show that the following 
expansion is valid for $1\leq k\leq g$:
\bea
&&
\partial_{A_k}\theta(\sum_{i=1}^k p_i+e|\Omega)={\tilde c}_k\partial_{A_{k-1}}\theta(\sum_{i=1}^{k-1} p_i+e|\Omega)z_k^{\lambda_k}+O(z_k^{\lambda_k+1}),
\non
\ena
where $z_k=z(p_k)$ and $c_k=\pm1$ is explicitly given (Theorem \ref{theta-AJimage}). 
The non-vanishing of the left hand side follows from (\ref{intro-1}).
This type of expansion was first pointed out in \cite{O1} in the case of hyperelliptic curves and $e=-\delta$ and was applied to addition formulae of the  fundamental sigma function .
 The results are extended to the case of  $(n,s)$ curves and $e=-\delta$ in \cite{NY,MP} and to that of telescopic curves and $e=-\delta$ in \cite{AN}.
 Here we extend the results to the case of an arbitrary Riemann surface and an arbitrary point $e$ on the theta divisor. 

The results on the expansions of the theta function above and the way to prove it implies 
an interesting extension and a refinement of Riemann's singularity theorem.

 Riemann's singularity theorem asserts that the multiplicity of $\theta(Z|\Omega)$ at $e$ is $m_0$ \cite{F,FK}.
In other words it says that any derivative of $\theta(Z|\Omega)$ of degree less than $m_0$ vanishes at $e$ and some derivative of degree $m_0$ does not vanish 
at $e$, where the degree signifies the degree as a differential operator. Here we should notice that 
the theorem tells nothing on which derivatives do not vanish in general. 

We derive the following properties of the theta function from those of $\tau$ functions and Schur functions:

\vskip5mm
\hskip5mm
(i) $\partial_I\theta(e|\Omega)=0$ for any $I=(i_1,...,i_m)$  if $i_1+\cdots+i_m<|\lambda|$.
\vskip3mm
\hskip5mm
(ii)  $\partial_I\theta(e|\Omega)=0$  for any $I=(i_1,...,i_m)$ if $m<m_0$. 
\vskip3mm
\hskip5mm
(iii) $\partial_{A_0}\theta(e|\Omega)\neq 0$.
\vskip5mm

The properties (ii) and (iii), in particular, implies Riemann's singularity theorem. Moreover we see that 
the $A_0$-derivative  gives the non-vanishing derivative of degree $m_0$ explicitly. The vanishing property 
(i) is a new vanishing property which does not follow from Riemann's singularity theorem.
Therefore (i)-(iii) give an extension and a refinement of Riemann's singularity theorem.

Finally this $A_0$-derivative  can be used to define an appropriate normalization in defining the sigma 
function such that the resulting function becomes modular invariant. In order to define sigma 
functions we need a certain bilinear meromorphic differential. The normalized bilinear differential  $\omega(p_1,p_2)=d_{p_1}d_{p_2}\log E(p_1,p_2)$, where $E(p_1,p_2)$ is the prime form,  plays a 
fundamental role in the theory of theta functions \cite{F}. However $\omega(p_1,p_2)$ depends on the choice 
of canonical homology basis. Klein modified $\omega$ so that it does not depend on the choice 
of canonical homology basis \cite{F, K}, which we call Kein form. Let $\widehat{\omega}(p_1,p_2)$ be a bilinear 
differential which is obtained from 
the Klein form by adding $\sum c_{ij} du_{w_i}du_{w_j}$, where $\{c_{ij}\}$ are independent of the choice of canonical homology basis and satisfy $c_{ij}=c_{ji}$. The sigma function associated with $(X,\{\alpha_i,\beta_i\},p_\infty, z, e, \{du_{w_i}\}, \widehat{\omega})$ is defined as follows. 

Let us write $e=\Omega\varepsilon'+\varepsilon''$ with $\varepsilon',\varepsilon''\in {\mathbb R}^g$
and set $\varepsilon={}^t(\varepsilon',\varepsilon'')$. Using Riemann's theta function with the characteristics 
$\varepsilon$ we set 
\bea
&&
C_e=\partial_{A_0}\theta[\varepsilon](0|\Omega),
\non
\ena
which does not vanish due to (iii) above.
Then we define the sigma function with the characteristics $\varepsilon$ by
\bea
&&
\sigma[\varepsilon](u)=C_{e}^{-1}\exp(\frac{1}{2}{}^tu\eta_1\omega_1^{-1} u)
\theta[\varepsilon]((2\omega_1)^{-1}u\,|\,\Omega),
\non
\ena
where $\eta_1$ is the period of certain second kind differentials which is computed from the $\alpha_i$-integral of $\widehat{\omega}$. The part without $C_e^{-1}$ of the right hand side, which we call the main part,  is already proposed in \cite{BEL1} without explicit construction of $\eta_1$. In \cite{KS} Korotkin and Shramchenko proposed to use Klein form to define $\eta_1$. They have shown that the main part,  multiplied by a certain theta constant which is apparently different from $C_e$, is invariant 
under the change of the canonical homology basis up to multiplication of $8N$-th root of unity, where 
$N$ is the number of non-singular even half periods.
We show that the sigma function normalized by $C_e$ is invariant under the action of $Sp(2g,{\mathbb Z})$
 on canonical homology basis.  We call this property the modular invariance of the sigma function.

There remain several fundamental problems to be solved. We have determined the initial term of the expansion of the theta function with respect to weight. In applications sometimes the initial term with respect to degree is necessary\cite{N3}. In \cite{BEL1} the minimal degree term is determined for a hyperelliptic curve with
$e$ being certain half periods as certain determinants.
In this paper we have determined the minimal degree term in the minimal weight term for arbitrary $(X,e)$. 
It is interesting to determine the full minimal degree term. To this end it is necessary to 
study higher weight terms in the expansion of $\tau$-function. The results in this direction
can be applied to the study on inversions of hyperelliptic integrals \cite{EHKKLP}.

The relation of Klein form  with the bilinear meromorphic differentials of $(n,s)$ curves, telescopic curves and 
others \cite{BEL3,BEL1,A,N1,KM,KMP}, which are constructed algebraically, should be clarified (see  \cite{EEE} for some examples). It is also  
interesting to determine  the explicit relation  between the normalization 
constants given in this paper and those in \cite{KS}. In the case of genus one the relation is given by 
the celebrated Jacobi's derivative formula.

In the case of  an $(n,s)$ curve the coefficients of the series expansion of the fundamental sigma function, which corresponds to $e=-\delta$, are polynomials of the coefficients of the defining equation 
of the curve \cite{N1,N2}. 
It is known that to any algebraic curve there exists a certain normal form of defining equations \cite{Miu}.
It is expected  that the coefficients of the series expansion of the fundamental sigma function of an algebraic curve can be expressed by the coefficients of defining equations of the curve as in the case of an $(n,s)$ curve. To this end we need to construct the (modified) Klein form algebraically using the defining equation of the curve. This construction is an independent interesting problem.

The paper is organized as follows.
In section 2 after the the review on divisors and line bundles on a Riemann surface we define the partition corresponding to a geometric data using gaps of line bundles. The properties of  the Schur functions corresponding to geometric data are studied in section 3.
 In section 4 the properties of the function which has a similar expansion to $\tau$-function of the KP-hierarchy is studied. Sato's theory of the KP-hierarchy is reviewed in section 5. In section 6 the point of UGM corresponding to an algebro-geometric solution of the KP-hierarchy is determined.  It is shown that the solution corresponding to 
 $(X,\{\alpha_i,\beta_i\},p_\infty,z,e)$ is in the cell $UGM^\lambda$ of UGM  with the partition $\lambda$ 
corresponding to this geometric data. This is 
the extension of the result in \cite{KNTY} to the non-generic case. The series expansions of the theta function are studied in section 7. We give examples of partitions corresponding to geometric data here. In the case of the hyperelliptic curve defined by an odd degree polynomial, the partition $\lambda$ corresponding to any data  of the form  $(X,\{\alpha_i,\beta_i\},\infty,e)$ are determined explicitly. In section 8 sigma functions with arbitrary real characteristics are defined and they are shown to be modular invariant.

%%%%%%%%%%%%%%%%%%%%%%%%%%%%%%%%%%%%%%%%%%%%%%%%%%%%%%%%%%%%%%%%%%%%%%%%%%%%%%%%%%%%%%%%%%%%%%%%%
\section{Geometric Data}
\subsection{Preliminaries}
Here we collect necessary facts on Riemann surfaces following mainly \cite{F}.

Let $X$ be a compact Riemann surface of genus $g$ and  $\{\alpha_i,\beta_i\}$ a canonical homology basis 
of $X$. Then the normalized basis $\{dv_i\}$ of holomorphic one forms and the normalized period matrix 
is determined:
\bea 
&& \int_{\alpha_j} dv_i=\delta_{ij},\qquad\qquad \Omega=\left(\int_{\beta_j} dv_i\right).
\non
\ena
 Riemann's theta function with characteristics $\varepsilon={}^t(\varepsilon',\varepsilon'')$, 
$\varepsilon', \varepsilon''\in {\mathbb R}^g$ is defined by 
\bea
&&
\theta[\varepsilon](z|\Omega)=\sum_{n\in {\mathbb Z}^g}\exp(\pi {}^t(n+\varepsilon')\Omega (n+\varepsilon')+2\pi i {}^t (n+\varepsilon')( z+\varepsilon'')),
\qquad
z={}^t(z_1,...,z_g).
\non
\ena
For $\varepsilon={}^t(0,0)$,  $\theta[\varepsilon](z|\Omega)$ is denoted by $\theta(z|\Omega)$.

Let $J(X)={\mathbb C}^g/({\mathbb Z}^g+\Omega {\mathbb Z}^g)$ be the Jacobian 
variety $X$.  By fixing a base point $p_\infty$ we have the Abel-Jacobi map 
:\bea
&&
I: X\rightarrow J(X), \qquad I(p)= \int_{p_\infty}^p dv,
\non
\ena
where $dv$ is the vector of normalize holomorphic one forms, $dv={}^t(dv_1,...,dv_g)$. 
The Jacobian $J(X)$ is isomorphic to the group of dvisor classes of degree zero by 
the Abel-Jacobi map.  We sometimes identify a divisor of degree zero with its Abel-Jacobi 
image in $J(X)$:
\bea
&&
\sum (p_i-q_i)=\sum\left(I(p_i)-I(q_i)\right).
\non
\ena
A choice $\{\alpha_i,\beta_i\}$ specifies the Riemann divisor $\Delta$. 
It is a divisor class of degree $g-1$ and satisfies $2\Delta=K_X$, where $K_X$ is the canonical 
divisor class of $X$.
Further if a point $p_\infty$ on $X$ is specified,  Riemann's constant 
$\delta$ is determined as an element of $J(X)$.  It is related to $\Delta$ by
\bea
&&
\Delta-(g-1)p_\infty=\delta.
\non
\ena

To each divisor $D$ is associated a holomorphic line bundle $L_D$ on $X$ (see \cite{F} for example). 
If $D$ is of degree zero, then 
\bea
&&
\frac{\theta\left(p-p_{\infty}-D-f\right)}{\theta\left(p-p_\infty-f\right)}
\non
\ena
is a meromorphic section of $L_D$, where $f$ is a generic point of 
${\mathbb C}^g$ such  that 
both denominator and numerator do not vanish identically.
 In terms of the transformation law,   if $D$ is represented by $c={}^t(c_1,...,c_g)\in {\mathbb C}^g$ in $J(X)$,
 a section $F$ of $L_{D}$ satisfies 
\bea 
&&
F(p+\alpha_j)=F(p),\quad F(p+\beta_j)=e^{2\pi i c_j} F(p).
\label{trf-rule}
\ena
In this case $L_D$ is also denoted by $L_c$.
A different choice of the representative $c$ of $D$ gives a holomorphically equivalent line bundle.
The holomorphic line bundle corresponding to a divisor of degree zero is called a flat line bundle 
since it admits a holomorphic flat connection. 

To each holomorphic line bundle there corresponds the sheaf of germs of holomorphic sections of it.
We shall introduce some notation related with this sheaf.

For two positive divisors $A$, $B$ and a holomorphic line bundle $L$ we denote by $L(B-A)$ the 
sheaf of germs of meromorphic sections of $L$ whose poles are at most at $B$ and whose zrors are 
at least at $A$. Let ${\cal O}$ denote the sheaf corresponding to the trivial line bundle $L_0$. 
Then  we have
\bea
&&
L_D\simeq {\cal O}(D).\qquad
L_D(B-A)\simeq {\cal O}(D+B-A),
\label{l-od}
\ena
as sheaves of ${\cal O}$ modules.

We set 
\bea
&&
h^0(L(B-A))=\dim H^0(X,L(B-A)).
\non
\ena
By (\ref{l-od}) we have 
\bea
&&
h^0(L_D(B-A))=h^0\bigl({\cal O}(D+B-A)\bigr).
\non
\ena
We denote the right hand side of this equation by $h^0(D+B-A)$.

For a point $p$ of $X$ let $L(\ast p)$ denote the sheaf of germs of meromorphic sections of $L$ which have 
a pole of any order at $p$.

%%%%%%%%%%%%%%%%%%%%%%%%%%%%%%%%%%%%%%%%%%%%%%%%%%%%%%%%%%%%%%%%%%%%%%%%%%%%%%%%%%%%%%%%%
\subsection{Gaps}

\begin{defn}
A non-negative integer $b$ is called a gap of a flat 
line bundle $L$ at $p_\infty$ if there does not exist a meromorphic section of $L$ which is holomorphic 
outside $p_\infty$ and has a pole of order $b$ at $p_\infty$. 
A non-negative integer which is not a gap is called a non-gap of $L$ at $p_\infty$.
\end{defn}

\begin{lemma}\label{lem-1}There are exactly $g$ gaps for any pair $(L,p_\infty)$ of a flat line bundle $L$ and 
a point $p_\infty$ of $X$.
\end{lemma}
\vskip2mm
\noindent
{\it Proof.} Let $L$ correspond to the divisor $D$. By the Riemann-Roch theorem we have
\bea
&&
h^0(D+np_\infty)-h^0(K_X-D-np_\infty)=1-g+n.
\label{eq-RR}
\ena

If $n\geq 2g-1$,
\bea
&&
h^0(K_X-D-np_\infty)=0,
\non
\ena
since the degree of $K_X-D-n$ is negative. In particular
\bea
&&
h^0(D+(2g-1)p_\infty)=g.
\non
\ena
Since we have $2g$ integers between $0$ and $2g-1$, it means that there are $g$ gaps in 
$\{0,1,...,2g-1\}$. It is obvious that there are no other gaps by (\ref{eq-RR}).
$\Box$

\subsection{Partition associated with a geometric data}

Let $e$ be a zero of the theta function $\theta(z\,|\,\Omega)$.
By Riemann's vanishing theorem $e$ can be written as 
\bea
&&
e=q_1+\cdots+q_{g-1}-\Delta,
\label{def-e}
\ena
for some $q_1,...,q_{g-1}\in X$. 

We consider the divisor $e+\delta$ of degree zero:
\bea
&&
e+\delta=q_1+\cdots+q_{g-1}-(g-1)p_\infty.
\label{def-L}
\ena

The proof of Lemma \ref{lem-1} shows that there are exactly $g$ gaps and $g$ non-gaps 
of $(L,p_\infty)$ 
in $\{0,1,...,2g-1\}$

We introduce two kinds of gaps and non-gaps simultaneously.
Let 
\bea
&&
b_1<b_2<\cdots<b_g,
\non
\\
&&
b_1^\ast<b_2^\ast<\cdots
\non
\ena
be gaps and non-gaps of  $L_{e+\delta}$ at $p_\infty$ and 
\bea
&&
w_1<w_2<\cdots<w_g,
\non
\\
&&
w_1^\ast<w_2^\ast<\cdots
\non
\ena
those of  $L_0$ at $p_\infty$.

\vskip2mm
\noindent
{\bf Remark} 
The sequence $(w_1,...,w_g)$ is the gap sequence 
of a Riemann surface $X$ at $p_\infty$ \cite{FK}. A point $p_\infty$ for which $(w_1,...,w_g)\neq 
(1,...,g)$ is called a Weierstrass point.  We have $w_1=1$ and $w_1^\ast=0$.
\vskip2mm

\begin{defn}\label{def-1} We define the partion $\lambda=(\lambda_1,...,\lambda_g)$ associated with
$(X, \{\alpha_i,\beta_i\}, p_\infty, e)$ by 
\bea
&&
\lambda=(b_g,...,b_1)-(g-1,...,1,0).
\label{lambda}
\ena
\end{defn}

In the next section we study the properties of  the Schur function corresponding to $\lambda$.

%%%%%%%%%%%%%%%%%%%%%%%%%%%%%%%%%%%%%%%%%%%%%%%%%%%%%%%%%%%%%%%%%%%%%%%%%%%%%%%%%%%%%%%%%%%
\section{Schur function of a Riemann surface}

\subsection{Dependence on variables}
The Schur function $s_\mu(t)$, $t=(t_1,t_2,t_3,..)$, \cite{Mac} corresponding to a partition $\mu=(\mu_1,...,\mu_l)$ is defined
by 
\bea
&&
s_\mu(t)=\det\left(p_{\mu_i-i+j}(t)\right)_{1\leq i,j\leq l},
\non
\ena
where $p_n(t)$ is the polynomial defined by 
\bea
&&
\exp\Bigl(\sum_{n=1}^\infty t_nk^n\Bigr)=\sum_{n=0}^\infty p_n(t) k^n,
\qquad
p_n(t)=0 \text{ for $n<0$}.
\non
\ena
We identify a partition $\mu=(\mu_1,...,\mu_l)$ with $(\mu_1,...,\mu_l,0^i)$ for any $i$. The Schur function $s_\mu(t)$ 
does not depend on the choice of $i$.

We define the weight and degree of the variable $t_i$ to be $i$ and $1$ respectively:
\bea
&&
\wt(t_i)=i,\qquad \deg(t_i)=1.
\non
\ena

With respcet to weight the Schur function $s_\mu(t)$ is a homogeneous polynomial with the 
weight $|\mu|=\mu_1+\cdots+\mu_l$, while it is not homogeneous with respect to degree in general.

By the definition $s_\mu(t)$ is a polynomial of $t_1,...,t_{\mu_1+l-1}$. However 
the Schur function corresponding to a geometric data depends on fewer number of variables.

\begin{prop} Let $\lambda$ be the partition associated with  
$(X, \{\alpha_i,\beta_i\}, p_\infty, e)$. Then $s_\lambda(t)$ is a polynomial of  $t_{w_1},...,t_{w_g}$.
\end{prop}
\vskip2mm
\noindent
{\it Proof.}  Notice that the space  $H^0(X,L_{e+\delta}(\ast p_\infty))$ is a 
$H^0(X,{\cal O}(\ast p_\infty))$-module. Therefore if we define  $N_1=\{w_i^\ast|\,i\geq 1\}$ and $N_2=\{b_i^\ast|\,i\geq 1\}$, then $N_1$ acts on $N_2$ by addition. Namely 
for any $i,j$ we have
\bea
&&
w_i^\ast+b_j^\ast\in N_2.
\non
\ena
It follows that, if  $b_j-w_i^\ast\geq0$ then it is a gap of $L$ at $p_\infty$. In fact if it is not the case, 
 $b_j-w_i^\ast\in N_2$. Then $b_j\in N_2+w_i^\ast \subset N_2$ which is absurd.
The proposition can be proved in a similar manner to  Proposition 2 of \cite{NY} using this property.
 $\Box$

\subsection{$a$-sequence}

\begin{defn}
For an integer $k$ such that $0\leq k\leq g-1$  we define the integer $m_k$ by
\bea
&&
m_k=h^0(q_1+\cdots+q_{g-1}-kp_\infty).
\non
\ena
\end{defn}

It is possible to describe $m_k$ in terms of gaps or non-gaps of $L_{e+\delta}$.

\begin{lemma}\label{mk}
We have 
\bea
&&
m_k=\sharp\{i\,|\, b_i^\ast <g-k\}=g-k-\sharp\{i\,|\, b_i<g-k\}.
\non
\ena
\end{lemma}
\vskip2mm
\noindent
{\it proof.} By the definition of $L_{e+\delta}$ we have 
\bea
&&
m_k=h^0\bigl( L_{e+\delta}((g-k-1)p_\infty)\bigr),
\non
\ena
which proves the first equation of the lemma.  Since gaps and non-gaps are complements to each other 
in $\{0,1,...,g-k-1\}$ the second equality follows.  $\Box$

In order to describe more detailed properties of $s_\lambda(t)$ we need

\begin{defn} For $0\leq k\leq g-1$ define the sequence $A_k=(a^{(k)}_1,...,a^{(k)}_{m_k})$ by
\bea
&&
A_k=(b_{g-k},b_{g-k-1},...,b_{g-k-m_k+1})-(b_1^\ast,...,b_{m_k}^\ast).
\non
\ena
The sequence $A_k$ is referred to as a-sequence.
\end{defn}

For a partition $\mu=(\mu_1,...,\mu_l)$ and $0\leq k\leq l-1$ we set 
\bea
&&
N_{\mu,k}=\sum_{i=k+1}^l \mu_i.
\non
\ena
In the case $k=0$,  $N_{\mu,0}=\sum_{i=1}^l \mu_i=|\mu|$ is the weight of $\mu$.

\begin{lemma}\label{lem-aseq} Suppose that $m_k>0$. Then 
\vskip2mm
\noindent
(i) $a^{(k)}_1>\cdots>a^{(k)}_{m_k}$.
\vskip2mm
\noindent
(ii)  $a^{(k)}_i\in \{w_j\}$ for any $i$.
\vskip2mm
\noindent
(iii) $\displaystyle{\sum_{i=1}^{m_k} a^{(k)}_i=N_{\lambda,k}}$. 
\end{lemma}
\vskip2mm
\noindent
{\it Proof.} (i) is obvious from the definition of $a^{(k)}_i$.
Let us prove (ii). We first show that $a^{(k)}_{m_k}>0$. By Lemma \ref{mk} we have 
$b_1^\ast<\cdots<b_{m_k}^\ast<g-k$. Since $\{b_i^\ast\}$ and $\{b_i\}$ are complement in 
the set of nonnegative integers to each other, 
\bea
&&
\{b_1^\ast,...,b_{m_k}^\ast\}\sqcup \{b_1,...,b_{g-k-m_k}\}=\{0,1,...,g-k-1\}.
\label{mk-division}
\ena
Thus $b_{m_k}^\ast<g-k\leq b_{g-k-m_k+1}$ and $a^{(k)}_{m_k}=b_{g-k-m_k+1}-b^\ast_{m_k}>0$. 
Now,  suppose that $a^{(k)}_i=b_{g-k+1-i}-b_i^\ast\notin \{w_j\}$. Since $a^{(k)}_i>0$, we have 
$a^{(k)}_i=w_j^\ast$ for some $j$. Then $b_{g-k+1-i}=b_i^\ast+w_j^\ast$ is a non-gap of $L_{e+\delta}$, which 
is impossible. Thus the assertion (ii) is proved.

\noindent
(iii): We have
\bea
\sum_{i=1}^{m_k}a^{(k)}_i&=&\sum_{i=g-k-m_k+1}^{g-k} b_i-\sum_{i=1}^{m_k} b_i^\ast
=\sum_{i=1}^{g-k} b_i-\sum_{i=1}^{g-k-1} i,
\label{sum-aki}
\ena
where we use (\ref{mk-division}). Since $\lambda_i=b_{g+1-i}-(g-i)$, the right hand side of 
 (\ref{sum-aki}) equals to $\lambda_{k+1}+\cdots+\lambda_g$.
 $\Box$

\subsection{Vanishing and non-vanishing}

We introduce the analogue of the Abel-Jacobi map for Schur function.

\begin{defn} Define $[x]$ by
\bea
&&
[x]=(x,\frac{x^2}{2},\frac{x^3}{3},...).
\non
\ena
\end{defn}

Using the a-sequence we can describe the properties of derivatives of $s_\lambda(t)$.

For a sequence $I=(i_1,...,i_r)$ of positive integers we set 
\bea
&&
\partial_{t,I}=\partial_{t_{i_1}}\cdots \partial_{t_{i_r}},\qquad
\partial_{t_i}=\frac{\partial}{\partial t_i}.
\non
\ena
In the following we sometimes use the expressions $\sum_{i=1}^k [x_i]$ and  $s_{(\mu_1,...,\mu_k)}(t)$ for 
$k=0$. 
They should be understood as 
\bea
&&
\sum_{i=1}^k [x_i]=0,\qquad s_{(\mu_1,...,\mu_k)}(t)=1.
\non
\ena

\begin{theorem}\label{th-1}
Suppose $m_k>0$. Let $\lambda$ be the partition defined by (\ref{lambda})
 and $\mu=(\mu_1,...,\mu_l)$ a partition satisfying $\mu_i=\lambda_i$ for 
$i\geq k+1$. Then
\bea
&&
\partial_{t,A_k}
s_\mu (\, \sum_{i=1}^k [x_i]\, )=
c_k s_{(\mu_1,...,\mu_k)}(\,\sum_{i=1}^k [x_i]\, ),
\non
\ena
where 
\bea
&&
c_k={\text sgn} \left(
\begin{array}{cccccc}
b_1^\ast&\cdots&b_{m_k}^\ast&b_{g-k-m_k}&\cdots&b_1\\
g-k-1&\cdots&\cdot&\cdot&\cdots&1\\
\end{array}
\right).
\label{signature}
\ena
\end{theorem}
\vskip2mm
\noindent
{\it Proof.} The theorem can be proved in a completely similar way to Theorem 1 in \cite{NY}.
 $\Box$
\vskip2mm

For two partitions $\mu=(\mu_1,...,\mu_l)$, $\nu=(\nu_1,...,\nu_{l'})$ we define $\mu\leq \nu$
 by the condition  $\mu_i\leq \nu_i$ for any $i$.

Then we have the following vanishing theorem for Schur functions.

\begin{theorem}\label{th-2}
Let $\lambda$ be the partition defined by (\ref{lambda}). Then 
\vskip2mm
\noindent
(i) Let $\mu$ be any partition. Then, for any sequence $I=(i_1,...,i_m)$, $m\geq 1$ satisfying 
$\sum_{j=1}^m i_j\neq |\mu|$ we have 
\bea
&&
\partial_{t,I}s_\mu(\,0\,)=0,
\label{der-0}
\ena
\vskip2mm
\noindent
(ii)  Let $\mu$ be a partition satisfying 
$\mu\geq \lambda$. If $m<m_0$ we have 
\bea
&&
\partial_{t,I}s_\mu(\,0\,)=0,
\label{der-1}
\ena
for any  and $I=(i_1,...,i_m)$.
\vskip2mm
\noindent
(iii) $\displaystyle{\partial_{t,A_0}
s_\lambda (\, 0\, )=c_0}$, where $c_0=\pm1$ is given by (\ref{signature}).
\end{theorem}
\vskip2mm
\noindent
{\it Proof.} 
Since $s_\mu(t)$ is weight-homogeneous with the weight $|\mu|$, (i) is obvious.
(iii) is the case of $k=0$ of Theorem \ref{th-1}. So let us prove (ii).

 If $\sum_{j=1}^m i_j\neq |\mu|$, the left hand side of (\ref{der-1}) vanishes by (i).
So we assume $\sum_{j=1}^m i_j=|\mu|=N_{\mu,0}$.

Let $[i_1,...,i_l]$ be the determinant of the $l\times l$ matrix whose $j$-th row is given by
\bea
&&
(p_{i_j-l+1}(t),...,p_{i_j-1}(t),p_{i_j}(t)).
\non
\ena
We write $[i_1,...,i_l](t)$ if it is necessary to indicate $t$. 
Let us define the strictly decreasing sequence $(b_l',...,b_1')$ corresponding to $\mu=(\mu_1,...,\mu_l)$ 
by
\bea
&&
(b_l',...,b_1')=(\mu_1,...,\mu_l)+(l-1,...,1,0).
\non
\ena
We take $l\geq g$ by inserting several $0$'s to the end of $\mu$ if necessary.
Then the condition $\mu\geq \lambda$ is 
\bea
&&
b_i'\geq b_{i-l+g}+l-g,\qquad l-g+1\leq i\leq l.
\non
\ena
With this notation we have
\bea
&&
s_\mu(t)=[b_l',...,b_1'].
\non
\ena
Since $\partial_{t,i} p_j(t)=p_{j-i}(t)$ and the derivative of the determinant by $\partial_{t,i}$ is the sum of the determinant whose $j$-th row is 
differentiated, we have
\bea
&&
\partial_{t,i}s_\mu(t)=\sum_{j=1}^l[b_l',...,b_j'-i,...,b_1'].
\non
\ena
Thus the left hand side of (\ref{der-1}) is written as a sum of the determinants of the form 
\bea
&&
[b_l'-r_l,...,b_1'-r_1](0).
\label{der-1-1}
\ena
If $r_j>0$ then it means that the $j$-th row is differentiated at least once.

We show that all terms (\ref{der-1-1}) appearing in the left hand side of (\ref{der-1}) vanish.

Suppose that there is a non-zero term  (\ref{der-1-1}). 
Since the number of the derivatives in the left side of (\ref{der-1}) is $m<m_0$, 
some row  among $m_0$ rows labeled by $l-m_0+1,..., l$  is not differentiated. 
We call this row the $j$-th row.
By (\ref{mk-division})  we have
\bea
&&
b_1<\cdots<b_{g-m_0}<g\leq b_{g-m_0+1}<\cdots<b_{g}.
\non
\ena
Thus 
\bea
&&
b_j'\geq b_{j-l+g}+l-g\geq l,
\non
\ena
since $g-m_0+1\leq j-l+g\leq g$ for $l-m_0+1\leq j\leq l$. By Lemma 2 (ii) in \cite {NY}
such a term is zero, which contradicts the assumption. 
$\Box$

\vskip2mm
\noindent
{\bf Remark} The assertions (ii) and (iii) of Theorem \ref{th-2} is an analogue of Riemann's singularity theorem for  Schur functions.

\subsection{Minimal degree term}

Assertions (ii)  and (iii) of Theorem \ref{th-2} mean that the term $t_{a^{(0)}_1}\cdots t_{a^{(0)}_{m_k}}$ is 
one of monomials  with the minimal degree which appear in $s_\lambda(t)$.
In fact it is possible to determine the minimal degree 
term of $s_\lambda(t)$. 

Let  $\mu=(\mu_1,...,\mu_l)$ be a partition and $L_\mu(t)$ the minimal degree term of $s_\mu(t)$:
\bea
&&
s_\mu(t)=L_\mu(t)+\text{higher degree terms}.
\non
\ena

Define $m_0(\mu)$ by 
\bea
&&
m_0(\mu)=l-\sharp\{\,i\,|\,b_i'<l\,\},\qquad (b_l',...,b_1')=(\mu_1,...,\mu_l)+(l-1,...,1,0).
\non
\ena
By Lemma \ref{mk} we have $m_0(\lambda)=m_0$.

\begin{prop}\label{minimal-wt-deg} 
We have
\bea
&&
L_\mu(t)=(-1)^{N_{\mu,m_0(\mu)}} \det(t_{\mu_i-i+j})_{1\leq i\leq m_0(\mu), j\neq l-b_1',...,l-b_{l-m_0(\mu)}'}.
\non
\ena
In particular the degree of $L_\mu(t)$ is $m_0(\mu)$.
\vskip2mm
\noindent
\end{prop}
\vskip2mm
\noindent
{\it Proof.} 
Since 
\bea
&&
p_n(t)=t_n+\text{higher degree terms}
\non
\ena
we have 
\bea
&&
s_\mu(t)=\det(t_{\mu_i-i+j})_{1\leq i,j\leq l} +\text{higher degree terms},
\label{smu-expand}
\ena
where we set $t_0=1$ and $t_i=0$ for $i<0$. 
Consider the determinant in the right hand side of (\ref{smu-expand}).
Consider $i$ with $b_i'<l$.
The $i$-th row from the bottom of the matrix $(t_{\mu_i-i+j})_{1\leq i,j\leq l}$ is 
\bea
&&
(0,...,0,t_0,t_1,...,t_{b_i'}).
\non
\ena
We first expand the determinant in the $l$-th row, which corresponds to $i=1$, and pick up 
the term containing $t_0$. 
It is the determinant of the matrix which is obtained by removing $l$-th row and 
$(l-b_1')$-th column times $(-1)^{l+(l-b_1')}$. We proceed similarly for $(l-1)$-th row, $(l-2)$-th row,
..., $(m_0(\mu)+1)$-th row. Then we get 
\bea
&& 
s_\mu(t)=(-1)^{N_{\mu,m_0(\mu)}}\det(t_{\mu_i-i+j})_{1\leq i\leq m_0(\mu), \,\,j\neq l-b_1',...,l-b_{l-m_0(\mu)}'}+\text{higher degree terms}.
\non
\ena
What we have to check is that the first term of the right hand side is not identically zero. Notice the 
monomial in $t_i$'s in the determinant which is obtained by taking the product of anti-diagonal components.
This monomial is unique among the $m_0(\mu)!$ terms in the expansion of the determinant and has $\pm 1$ as its coefficient.  $\Box$
\vskip5mm

\noindent
{\bf Example} 
Consider a hyperelliptic curve of genus $g$ defined by $y^2=f(x)$, where $f(x)$ is a polynomial of degree $2g+1$ without multiple zeros and 
take $p_\infty =\infty$, $e=-\delta$. Then  $\lambda=(g,g-1,...,1)$.
In this case $m_0=[\frac{g+1}{2}]$, $N_{\lambda,m_0}=(1/2)(g-m_0)(g+1-m_0)$ and 
\vskip3mm
\hskip30mm
$L_\lambda(t)=(-1)^{N_{\lambda,m_0}}\det(t_{2k+1-2i+2j})_{1\leq i,j\leq k}$ for $g=2k$,
\vskip3mm
\hskip30mm
$L_\lambda(t)=(-1)^{N_{\lambda,m_0}}\det(t_{2k+1-2i+2j})_{1\leq i,j\leq k+1}$ for $g=2k+1$.
\vskip2mm
If we introduce the variables with different indices by
\bea
&&
(u_1,u_2,...,u_g)=(t_{2g-1},t_{2g-3},...,t_1),
\non
\ena
then 
\bea
&&
L_\lambda(t)=(-1)^{(1/2)g(g+1)+gm_0}
\left|\begin{array}{cccc}
u_1&u_2&\ldots&u_{m_0}\\
u_2&u_3&\ldots&u_{m_0+1}\\
\vdots&\quad&\quad&\vdots\\
u_{m_0}&u_{m_0+1}&\ldots&u_{2m_0-1}
\end{array}
\right|,
\non
\ena
which is precisely the Hankel determinant formula for the minimal degree term of the series expansion of 
 the hyperelliptic theta function derived in \cite{BEL1,B}.

%%%%%%%%%%%%%%%%%%%%%%%%%%%%%%%%%%%%%%%%%%%%%%%%%%%%%%%%%%%%%%%%%%%%%%%%%%%%%%%%%%%%%%%%%%%
\section{Tau function}
\subsection{Expansion on Abel-Jacobi images}
Let $\lambda$ be the partition (\ref{lambda}) associated with $(X, \{\alpha_i,\beta_i\}, p_\infty, e)$.
In this section we consider an arbitrary function of the form 
\bea
&&
\tau(t)=s_\lambda(t)+\sum_{\lambda<\mu}\xi_\mu s_\mu(t).
\non
\ena

For $1\leq k\leq g$ we set 
\bea
&&
\tau^{(k)}(t)=s_{(\lambda_1,...,\lambda_k)}(t)+\sum_\mu \xi_\mu s_{(\mu_1,...,\mu_k)}(t),
\non
\ena
where the sum in the right hand side runs over partitions $\mu=(\mu_1,...,\mu_g)$ satisfying the conditions 
$\lambda<\mu$, $\mu_i=\lambda_i$ for $k+1\leq i\leq g$. We set $\tau^{(0)}(t)=1$.

\begin{theorem}\label{tau-AYimage} Suppose that $m_k>0$. Let $c_k$ be given by (\ref{signature}). Then 
\vskip2mm
\noindent
(i) $\displaystyle{\partial_{t,A_k}\tau\left(\sum_{i=1}^k [x_i]\right)
=c_k \tau^{(k)}\left(\sum_{i=1}^k [x_i]\right)}$.
\vskip2mm
\noindent
(ii)  $\displaystyle{\tau^{(k)}\left(\sum_{i=1}^k [x_i]\right)=\tau^{(k-1)}\left(\sum_{i=1}^{k-1} [x_i]\right)x_k^{\lambda_k}+O(x_k^{\lambda_k+1}).}$
\vskip2mm
\noindent
(iii) $\displaystyle{\partial_{t,A_k}\tau\left(\sum_{i=1}^k [x_i]\right)
=\frac{c_k}{c_{k-1}}\partial_{t,A_{k-1}}\tau\left(\sum_{i=1}^{k-1} [x_i]\right)
x_k^{\lambda_k}+O(x_k^{\lambda_k+1}).}$
\end{theorem}
\vskip2mm
\noindent
{\it Proof.} The proof of the theorem is similar to that of Theorem 5 in \cite{NY}. For the sake of 
completeness we give a proof here. 

Let $\mu=(\mu_1,...,\mu_l)$ be a partition satisfying $\lambda\leq \mu$. Here $l$ is not necessarily the 
length of $\mu$. 
By (iii) of Lemma \ref{lem-aseq} we have  
\bea
&&
\sum_{i=1}^{m_k} a^{(k)}_i=\sum_{i=k+1}^{g} \lambda_i\leq \sum_{i=k+1}^{l} \mu_i=N_{\mu,k}.
\non
\ena

In case the last inequality is a strict inequality 
\bea
&&
\partial_{t,A_k}s_\mu\left(\sum_{i=1}^k [x_i]\right)=0
\label{der-smu}
\ena
by Proposition 3 in \cite{NY}. Thus, if the left hand side of (\ref{der-smu}) is not zero we have 
\bea
&&
\sum_{i=k+1}^{g} \lambda_i= \sum_{i=k+1}^{l} \mu_i
\non
\ena
which implies 
\bea
&&
\lambda_i=\mu_i, \quad k+1\leq i\leq g,
\non
\ena
and $\mu_i=0$ for $i>g$, since $\lambda\leq \mu$. Then we have, by Theorem \ref{th-1},
\bea
&&
\partial_{t,A_k}s_\mu\left(\sum_{i=1}^k [x_i]\right)=
c_k s_{(\mu_1,...,\mu_k)}\left(\sum_{i=1}^k [x_i]\right).
\non
\ena
The assertion (i) follows from this. The assertion (ii) is already proved in (ii) of Theorem 5 in \cite{NY}, since no specialty of the partition $\lambda$ associated with an $(n,s)$ curve is used there.  (iii) follows from (i) and (ii). $\Box$
\vskip2mm

\subsection{Vanishing and non-vanishing}
The following vanishing theorem for the function $\tau$ is valid.
\begin{theorem}\label{Rvanishing-tau} 
(i) For any $I=(i_1,...,i_m)$, $m\geq 1$, satisfying $\sum_{j=1}^m i_j<|\lambda|$ we have 
\bea
&&
\partial_{t,I}\tau(0)=0.
\non
\ena
\vskip2mm
\noindent
(ii) If $m<m_0$
\bea
&&
\partial_{t,I}\tau(0)=0,
\non
\ena
for any $I=(i_1,...,i_m)$.
\vskip2mm
\noindent
(iii)  $\displaystyle{\partial_{t,A_0}\tau(0)=c_0}$, where $c_0=\pm1$ 
is given by (\ref{signature}).
\end{theorem}
\vskip2mm
\noindent
{\it Proof.} The assertions of the theorem follow from (ii), (iii) of Theorem \ref{th-2} and the 
definition of the function $\tau(t)$. $\Box$
\vskip2mm

Notice that this theorem is an analogue of  Riemann's singularity theorem 
for the function $\tau(t)$.

%%%%%%%%%%%%%%%%%%%%%%%%%%%%%%%%%%%%%%%%%%%%%%%%%%%%%%%%%%%%%%%%%%%%%%%%%%%%%%%%%%%%%%
\section{Sato's theory on soliton equations}
In this section we  review  Sato's theory of the KP-hierarchy\cite{SS} which makes a one to one 
correspondence between solutions of the KP-hierarchy and points of an infinite dimensional 
Grassmann manifold, called the universal Grassmann manifold (UGM).

%%%%%%%%%%%%%%%%%%%%%%%%%%%%%%%%%%%%%%%%%%%%%%%%%%%%%%%%%%%%%%%%%%%%%%%%%%%%%%%%%%%%%
\subsection{The KP-hierarchy}
The KP-hierarchy is the infinite system of differential equations for $\tau(t)$ given by 
\bea
&&
\int \tau(t-s-[k^{-1}])\tau(t+s+[k^{-1}])e^{-2\sum_{i=1}^\infty s_ik^i} dk=0
\label{bilinear-eq}
\ena
where $t={}^t(t_1,t_2,,,,)$, $s={}^t(s_1,s_2,...)$ and the integral signifies taking residue at $k=\infty$ \cite{DJKM}.
Namely we formally expand the integrand in the series of $k$ and $y$ and equate the coefficient
of $k^{-1}s_1^{\gamma_1}s_2^{\gamma_2}\cdots$ to zero. Then we get an infinite number of 
differential equations for $\tau(t)$. 
 In particular taking the coefficient of $k^{-1}s_3$ we get 
the bilinear form of the Kadomtsev-Petviashvili (KP) equation:
\bea
&&
(D_1^4+3D_2^2-4D_1D_3)\tau(t)\cdot \tau(t)=0,
\non
\ena
where $D_i$ is the Hirota derivative defined by
\bea
&&
\tau(t+s)\tau(t-s)=\sum 
(D_1^{\gamma_1}D_2^{\gamma_2}\cdots)\tau(t)\cdot \tau(t)\,\,\,
\frac{s_1^{\gamma_1}s_2^{\gamma_2}\cdots}{\gamma_1!\gamma_2!\cdots}.
\ena

The initial value problem of the KP-hierarchy is uniquely solvable and the 
set of the initial values forms the infinite dimensional Grassmann manifold UGM.

%%%%%%%%%%%%%%%%%%%%%%%%%%%%%%%%%%%%%%%%%%%%%%%%%%%%%%%%%%%%%%%%%%%%%%%%%%%%%%%%%%%%
\subsection{UGM}
Let us give the definition of  UGM \cite{SS}.
To this end we consider the ring of microdifferential operators with the coefficients in the 
formal power series ring $R={\mathbb C}[[x]]$ in one variable $x$. Namely ${\cal E}_R$ consists 
of  all expressions of the form 
\bea
&&
a_n(x)\partial^n+a_{n-1}(x)\partial^{n-1}+\cdots,\qquad n\in{\mathbb Z},\quad a_i(x)\in R,
\non
\ena
where $\partial=\partial/\partial x$. Using 
\bea
&&
a\partial^n=\sum_{i=0}^\infty  (-1)^i\bc{n}{i}\partial^{n-i}a^{(i)}, \qquad
a^{(i)}=\frac{d^i a}{d x^i},
\non
\ena
elements of  ${\cal E}_R$ can equally be rewritten  in the form 
\bea
&&
\sum_{i\leq n} \partial^i b_i, \qquad n\in {\mathbb Z}, \qquad b_i\in R.
\label{der-left}
\ena

Next we intoduce the left ${\cal E}_R$-module $V$ by 
\bea
&&
V={\cal E}_R/{\cal E}_R x.
\non
\ena
The expression (\ref{der-left}) implies that $V$ is isomorphic to the space of microdifferential 
operators with constant coefficients:
\bea
&&
V\simeq {\mathbb C}((\partial^{-1})).
\label{const-coeff}
\ena
Then we see that $V$ has the decomposition of the form
\bea
&&
V=V_\phi\oplus V_0, \qquad
V_\phi={\mathbb C}[\partial],\qquad 
V_0={\mathbb C}[[\partial^{-1}]]\partial^{-1}.
\non
\ena
Let us define the element $e_i$ of $V$ by 
\bea
&&
e_i=\partial^{-i-1}\quad \text{mod.}  \,{\cal E}_R x,\qquad i\in {\mathbb Z}.
\non
\ena
Then 
\bea
&&
V_\phi=\oplus_{i=-\infty}^{-1} {\mathbb C} e_i,\qquad
V_0=\prod_{i=0}^\infty {\mathbb C} e_i.
\non
\ena
The action of ${\cal E}_R$ on $V$ is given by 
\bea
&&
\partial^{\pm1} e_i=e_{i\mp1},\qquad x e_i=(i+1)e_{i+1}.
\non
\ena

UGM is defined as the set of subspaces of $V$ which are comparable with $V_\phi$.
To be precise we need some notation. For a subspace $U$ of $V$ 
we denote $\pi_U$ the map
\bea
&&
\pi_U:U\longrightarrow V/V_0\simeq V_\phi,
\non
\ena
which is obtained as the composition of the inclusion $U\hookrightarrow V$ and the natural 
projection $V\rightarrow V/V_0$.

\begin{defn} The universal Grassmann manifold UGM is the set of subspaces 
$U\subset V$ such that $\text{Ker}\,\pi_U$ and $\text{Coker}\, \pi_U$ are of finite dimension and 
satisfy
\bea
&&
\dim(\text{Ker}\,\pi_U)-\dim(\text{Coker}\, \pi_U)=0.
\non
\ena
\end{defn}

A point $U$ of UGM is specified by giving a frame of $U$, which is an ordered basis of $U$.

To each point $U$ of UGM there exists the unique sequence of integers $\rho=(\rho(i))_{i<0}$ and 
the unique frame  $\bxi=(\xi_j)_{j<0}$, $\xi_j=\sum_{i\in{\mathbb Z} }\xi_{ij} e_i$,  of $U$ such that 
\bea
&&
\rho(-1)>\rho(-2)>\cdots, \qquad \rho(i)=i \text{ for $i<<0$},
\label{d-seq}
\\
&&
\xi_{ij}=\left\{
\begin{array}{ll}
0&\quad i<\rho(j) \text{ or } i=\rho(j') \text{ for some $j'>j$}\\
1&\quad i=\rho(j). 
\end{array}
\right.
\non
\ena
This frame  $\bxi$ is called the normalized frame of $U$. 

The dimensions of $\text{Ker}\,\pi_U$ and $\text{Coker}\,\pi_U$ are expressed 
by $\rho$:
\bea
&&
\dim(\text{Ker}\,\pi_U)=\sharp\{i\,|\,\rho(i)\geq 0\,\},
\qquad 
\dim(\text{Coker}\,\pi_U)=\sharp\{i<0\,|\,i \notin \{\rho(j)\}\,\}.
\non
\ena

In general for a sequence $\rho=(\rho(i))_{i<0}$ satisfying the condition (\ref{d-seq}) define the partition   $\lambda_\rho$ by 
\bea
&&
\lambda_\rho=(\lambda_{\rho,1},\lambda_{\rho,2},...)=(\rho(-1),\rho(-2),...)+(1,2,...).
\label{lambda-rho}
\ena
The partition $\lambda=(0,0,...)$ corresponding to $\rho=(-1,-2,...)$ is denoted by $\phi$.

Conversely for any partition $\lambda=(\lambda_1,...,\lambda_l)$ one can construct 
$\rho$ satisfying (\ref{d-seq}) using (\ref{lambda-rho}), where we set $\lambda_i=0$ for $i>l$.

For a partition $\lambda$ let $UGM^\lambda$ be the set of points $U$ of UGM such that 
the partition associated with $U$ is $\lambda$. Then UGM has the decomposition 
\bea
&&
UGM=\sqcup_\lambda UGM^\lambda,
\non
\ena
where $\lambda$ runs over all partitions.

%%%%%%%%%%%%%%%%%%%%%%%%%%%%%%%%%%%%%%%%%%%%%%%%%%%%%%%%%%%%%%%%%%%%%%%%%%%%%%%%%%%55
\subsection{Fundamental theorems of Sato's theory}
Let $U$ be a point of UGM, $\bxi=(\xi_j)_{j<0}$ the normalized frame of $U$ and 
$\mu$ an arbitrary partition. Let us write $\mu=\lambda_\rho$ with $\rho$ 
 satisfying (\ref{d-seq}). We define the Pl$\ddot{u}$cker 
coordinate $\xi_{\mu}$ of $U$ associated with $\mu$ 
by 
\bea
&&
\xi_{\mu}=\det\left(\xi_{\rho(i),j}\right)_{i,j<0}.
\non
\ena

If $U\in UGM^\lambda$ the Pl$\ddot{u}$cker coordinates satisfy 
\bea
&&
\xi_\mu=\left\{
\begin{array}{ll}
0&\quad \hbox{unless $\mu\geq \lambda$}\\
1&\quad \mu=\lambda.
\end{array}
\right.
\label{Plucker}
\ena

\begin{defn}
Define 
the tau function corresponding to $U$ by
\bea
&&
\tau(t;U)=\sum_{\mu}\xi_\mu s_\mu(t).
\label{exp-tau}
\ena
\end{defn}

\begin{theorem}\cite{SS,SN}\label{UGM-tau}
For any $U$,  $\tau(t;U)$  is a solution of the
KP-hierarchy. Conversely any formal power series solution $\tau(t)$
of the KP-hierarchy  there exists a unique point $U$ of UGM such that $\tau(t)=c \tau(t;U)$ 
for some constant $c$.
\end{theorem}

 Notice that, for a point $U$ of  $UGM^\lambda$, we have 
\bea
&&
\tau(t;U)=s_\lambda(t)+\sum_{\lambda< \mu}\xi_\mu s_\mu(t)
\label{exp-tau-lambda}
\ena
due to (\ref{Plucker}).

Now we explain how to recover $U$ from $\tau(t)$.

Let $K={\mathbb C}((x))$ be the field of formal Laurent series in $x$ and 
${\cal E}_K=K((\partial^{-1}))$ the ring of microdifferential operators with the coefficients in $K$.

\begin{defn}\label{pseudo-reg}
Let ${\cal W}$ be the set of $W$ in ${\cal E}_K$ of the form
\bea
&&
W=\sum_{i\leq 0}w_i \partial^i,
\quad
w_0=1,
\non
\ena
which satisfies the condition 
\bea
&&
x^mW, \quad W^{-1}x^m\in {\cal E}_R,
\label{quasi-regular}
\ena
for some non-negative integer $m$.
\end{defn}

Then 

\begin{theorem}\cite{SS,SN}\label{W-UGM}
Let $W$ be an element of ${\cal W}$ and $m$ an integer as in  (\ref{quasi-regular}). 
Then 
\bea
&&
\gamma(W)=W^{-1}x^mV^\phi
\non
\ena
defines a bijection $\gamma: {\cal W}\rightarrow UGM$.
\end{theorem}

We remark that the map $\gamma$ does not depend on the choice of $m$, since $x:V_\phi\rightarrow V_\phi$ is a surjection.

Given a formal power series  solution of the KP-hierarchy $\tau(t)\neq 0$ 
the wave function $\bar{\Psi}(t;z)$ and the adjoint wave function  $\Psi(t,z)$ are defined by
\bea
&&
\bar{\Psi}(t;z)=\frac{\tau(t-[z])}{\tau(t)}\exp(\sum_{i=1}^\infty t_iz^{-i}).
\non
\\
&&
\Psi(t;z)=\frac{\tau(t+[z])}{\tau(t)}\exp(-\sum_{i=1}^\infty t_iz^{-i}),
\non
\ena
Let 
\bea
&&
\frac{\tau(t-[z])}{\tau(t)}=\sum_{i=0}^\infty w_i z^i,
\qquad\qquad
W=\sum_{i=0}^\infty w_i \partial^{-i}.
\non
\ena
Then 
\bea
&&
\bar{\Psi}(t,z)=W\exp(\sum_{i=1}^\infty t_iz^{-i}),
\non
\ena
where we set $t_1=x$.
The equation (\ref{bilinear-eq}) implies that $\Psi$ can be written as
\bea
&&
\Psi(t,z)=(W^\ast)^{-1}\exp(-\sum_{i=1}^\infty t_iz^{-i}).
\non
\ena
where $P^\ast=\sum (-\partial)^i a_i(x)$ is 
the formal adjoint of $P=\sum a_i(x) \partial^i$.
We have $(P^\ast)^{-1}=(P^{-1})^\ast$ 
for an invertible $P\in {\cal E}_K$ \cite{DJKM}.

If $\tau(t)$ is not identically $0$,  $\tau(x,0,0,...)$ 
is not identically zero \cite{SW} (see Lemma 4 in \cite{N2}). 
Let $m_0$ be the order of zeros of $\tau(x,0,0...)$ at $x=0$ 
and $m\geq m_0$. Obviously we have 
\bea
&&
x^m W(x,0,...), \quad W(x,0,...)^{-1}x^m\in {\cal E}_R 
\non
\ena
which implies $W\in {\cal W}$.
Then 

\begin{theorem}\label{UGMpt-tau}\cite{SN,SS} There is a constant $C$ such that
\bea
&&
C\tau(t)=\tau\left(t;\gamma\left(W(x,0,...)\right)\right).
\non
\ena
\end{theorem}

The image $\gamma\left(W(x,0,...)\right)$ can be computed from $\Psi( t;z)$
using the following proposition \cite{N2} (see also  \cite{Shi2}\cite{KNTY}).

\begin{prop}\label{UGM-psi}
Let
\bea
&&
x^m\Psi(x,0,...;z)=\sum_{i=0}^\infty \Psi_i(z)\frac{x^i}{i!}.
\label{expansion-psi}
\ena
Then we have, for $i\geq 0$, 
\bea
&&
W(x,0,...)^{-1}x^me_{-1-i}=(-1)^i\Psi_i(\partial^{-1})e_{-1}.
\non
\ena
In particular 
\bea
&&
\gamma(W(x,0,...))=\mbox{Span}_{\mathbb C}\{\Psi_i(\partial^{-1})e_{-1}\,|\,
i\geq 0\,\},
\non
\ena
where $\mbox{Span}_{\mathbb C}\{\cdots\}$ signifies the vector space spanned
by $\{\cdots\}$.
\end{prop}

%%%%%%%%%%%%%%%%%%%%%%%%%%%%%%%%%%%%%%%%%%%%%%%%%%%%%%%%%%%%%%%%%%%%%%%%%%%%%%%%%%%%%
\section{Algebro-geometric solution}
In this section we determine the point of UGM corresponding to a theta function solution of the KP-hierarchy.
We assume that a data $(X,\{\alpha_i,\beta_i\},p_\infty,z)$ is given, where $(X,\{\alpha_i,\beta_i\},p_\infty)$
is as before and $z$ is a local coordinate at $p_\infty$.
\subsection{Prime form}

Let us first recall the prime form \cite{F}.
It is known that there exists a non-singular odd half period $e'\in J(X)$. 
We write $e'$ as 
\bea
&&
e'=q_1'+\cdots+q_{g-1}'-\Delta,\hskip10mm q_i'\in X.
\non
\ena
Let $\varepsilon_0={}^t(\varepsilon_0',\varepsilon_0'')$, 
$\varepsilon_0', \varepsilon_0''\in {\mathbb R}^g$ be the characteristics of $e'$:
\bea
&&
e'=\Omega \varepsilon_0'+\varepsilon_0''.
\non
\ena
Then the zero divisor of the holomorphic one form 
\bea
&&
\sum_{i=1}^g\frac{\partial\theta[\varepsilon_0]}{\partial z_i}(0) dv_i
\non
\ena
is $2\sum_{i=1}^{g-1}q_i'$.  Since $e'$ is non-singular, there exists 
a unique, up to constant multiples, holomorphic section of $L_{e'}\otimes L_\Delta$
which vanishes exactly at $q_1'+\cdots +q_{g-1}'$. 

Let $h_{\varepsilon_0}$ be a holomorphic 
section of  $L_{e'}\otimes L_\Delta$ such that 
\bea
&&
h_{\varepsilon_0}^2=\sum_{i=1}^g\frac{\partial\theta[\varepsilon_0]}{\partial z_i}(0) dv_i.
\non
\ena
Then the prime form $E(p_1,p_2)$ is defined by 
\bea
&&
E(p_1,p_2)=\frac{\theta[\varepsilon_0](\int_{p_1}^{p_2}dv)}{h_{\varepsilon_0}(p_1)h_{\varepsilon_0}(p_2)}.
\label{prime-form}
\ena
Let $\pi_i$ be the projection of $X\times X$ to the $i$-th component $X$ and $I_{21}$ the map 
from $X\times X$ to $J(X)$ defined by $I_{21}(p_1,p_2)=I(p_2)-I(p_1)$. We denote by $\Theta$ the holomorphic line bundle on $J(X)$ of which $\theta(z|\Omega)$ is a holomorphic section.

Then the prime form is a holomorphic section of 
$\pi_1^\ast L_\Delta^{-1}\otimes \pi_2^\ast L_\Delta^{-1}\otimes I_{21}^\ast(\Theta)$.
The prime form $E(p_1,p_2)$ is skew symmetric in $(p_1,p_2)$ and vanishes at the diagonal $\{(p,p)|p\in X\}$ 
 to the first order.
For other properties see \cite{F}.

We need the object which is obtained from the prime form by restricting one of the variables
 to a point.  Using the local coordinate $z$  around $p_\infty$  let us write 
\bea
&&
E(p_1,p_2)=\frac{E(z_1,z_2)}{\sqrt{dz_1}\sqrt{dz_2}},
\non
\ena
where $z_i=z(p_i)$.
Then we set 
\bea
&&
E(p,p_\infty)=\frac{E(z,0)}{\sqrt{dz}},
\label{prime-one}
\ena
which is a constant multiple of 
\bea
&&
\frac{ \theta[\varepsilon_o] ( I(p) \,|\,\Omega) }{ \sqrt{ h_{\varepsilon_o}(p) } }.
\non
\ena
The proportional constant is determined so that the expansion of $E(p,p_\infty)\sqrt{dz}$ at $p_\infty$ 
 has the form $-z+O(z^2)$. So it depends on the choice of $z$.
It is  a section of the line bundle $L_{\Delta}^{-1}\otimes I^\ast\Theta$.
The holomorphic line bundle  $I^\ast \Theta$ on $X$ is described by the transformation rule
\bea
f(p+\alpha_j)&=&f(p),
\non
\\
f(p+\beta_j)&=&e^{-\pi i \Omega_{jj}-2\pi i(\int_{p_\infty}^p dv_j)}f(p).
\label{trf-rule-Itheta}
\ena

%%%%%%%%%%%%%%%%%%%%%%%%%%%%%%%%%%%%%%%%%%%%%%%%%%%%%%%%%%%%%%%%%%%%%%%%%%%%%%%%%%%%
\subsection{Theta function solution}

Let 
\bea
&&
\omega(p_1,p_2)=d_{p_1}d_{p_2}\log E(p_1,p_2),
\non
\ena
be the fundamental normalized differential of the second kind \cite{F}. 
Using the local coordinate $z$ we expand $\omega(p_1,p_2)$ and $dv_i$ as
as 
\bea
\omega(p_1,p_2)&=&\left(\frac{1}{(z_1-z_2)^2}+\sum_{i,j=1}^{\infty}q_{ij}z_1^{i-1}z_2^{j-1}\right)dz_1dz_2,
\label{bilinear-omega}
\\
dv_i&=&\left(\sum_{j=1}^\infty a_{ij} z^{j-1}\right)dz,
\non
\ena
where $z_i=z(p_i)$. 
Set 
\bea
&&
q(t)=\sum_{i,j=1}^\infty q_{ij} t_i t_j,\qquad
A=(a_{ij})_{1\leq i\leq g, 1\leq j}.
\non
\ena
It is well known that 
\bea
&&
\tau(t)=e^{\frac{1}{2}q(t)}\theta(At+e\,|\,\Omega\,)
\label{theta-sol-KP}
\ena
is a solution of the KP-hierarchy (\ref{bilinear-eq}) for any $e={}^t(e_1,...,e_g)\in {\mathbb C}^g$ 
\cite{Kr, DJKM, SW, Shi, KNTY}.

%%%%%%%%%%%%%%%%%%%%%%%%%%%%%%%%%%%%%%%%%%%%%%%%%%%%%%%%%%%%%%%%%%%%%%%%%%%%%%%%%%%%%
\subsection{The point of UGM}
Let us determine the point of UGM corresponding to (\ref{theta-sol-KP}). 
To this end we compute the adjoint wave function associated with $\tau(t)$.

Let $dr_n$, $n\geq 1$ be the meromorphic one form with a pole only at $p_\infty$ of order $n+1$
which satisfies the following conditions:
\bea
&&
dr_n=d\left(\frac{1}{z^n}+O(z)\right)\qquad \text{near $p_\infty$},
\non
\\
&&
\int_{\alpha_i} dr_n=0 \text{ for any $i$}.
\non
\ena

Then we have \cite{Kr, KNTY,N2}
\bea
&&
z^{-1}\Psi(t;z)\sqrt{dz}=\frac{1}{E(p,p_\infty)}\frac{\theta(I(p)+At+e)}{\theta(At+e)}
\exp\Bigl(-\sum_{n=1}^\infty t_n\int^p dr_n\Bigr),
\label{theta-wave}
\ena
where the integral is normalized as 
\bea
&&
\lim_{p\rightarrow p_\infty}\left(\int^p dr_n-\frac{1}{z^n}\right)=0.
\non
\ena
Notice that
\bea
&&
\theta(I(p)+At+e)\exp\Bigl(-\sum_{n=1}^\infty t_n\int^p dr_n\Bigr)
\label{part-psi}
\ena
is a section of the holomorphic line bundle $I^\ast (\Theta)\otimes L_{-e}$ on $X$ and 
 specified by the transformation rule (\ref{trf-rule-delta-c}).
Therefore   $z^{-1}\Psi(t;z)\sqrt{dz}$ can be considered a section of the line bundle $L_{\Delta}\otimes L_{-e}$.

Now we define the map
\bea
&&
\iota: H^0\left(X,(L_\Delta\otimes L_{-e})(\ast p_\infty)\right)\longrightarrow V,
\non
\ena
in the following way. 

First we specify the local trivialization of $L_\Delta\otimes L_{-e}$ as in (\ref{theta-wave}).
Namely a section of this bundle is written as $E(p,p_\infty)^{-1}$ times a section of 
$L_{-e}\otimes I^\ast\Theta$ and the latter
 is realized as a multiplicative functions on $X$ which obeys the transformation rule  given by
\bea
f(p+\alpha_j)&=&f(p),
\non
\\
f(p+\beta_j)&=&e^{-\pi i\Omega_{jj}-2\pi i(\int_{p_\infty}^p dv_j+e_j)}f(p).
\label{trf-rule-delta-c}
\ena

Take an element $\varphi(p)$ of $H^0\left(X,(L_\Delta\otimes L_{-e})(\ast p_\infty)\right)$ and 
expand it around $p_\infty$ in $z$:
\bea
&&
\varphi(p)=\sum_{-\infty<<n<\infty} c_nz^n \sqrt{dz}.
\non
\ena
Then we set 
\bea
&&
\iota(\varphi)=\sum_{-\infty<<n<\infty} c_n e_n\in V.
\label{def-iota}
\ena

Define the subspace $U$ of $V$ as the image of $\iota$:
\bea
&&
U=\iota\left(H^0\left(X,(L_\Delta\otimes L_{-e})(\ast p_\infty)\right)\right).
\label{def-U}
\ena
Let $\lambda$ be the partition defined by  (\ref{lambda}) if $\theta(e|\Omega)=0$ and $e$ is given 
by (\ref{def-e}). We set $\lambda=\phi$ if $\theta(e|\Omega)\neq 0$. Then 

\begin{theorem}\label{U-UGM}The subspace $U$ is a point of $UGM^\lambda$.
\end{theorem}
\vskip2mm
\noindent
{\it Proof.} For $e$ satisfying $\theta(e|\Omega)\neq 0$ the theorem is proved 
in \cite{KNTY}.  Let us prove the theorem in case $\theta(e|\Omega)=0$. 

Writing $e$ as in  (\ref{def-e})  we have
\bea
L_{\Delta}\otimes L_{-e}&\simeq& K_X(-q_1-\cdots-q_{g-1}),
\label{delta-e-}
\\
L_{\Delta}\otimes L_{e}&\simeq&{\cal O}(q_1+\cdots+q_{g-1}).
\label{delta-e}
\ena
Then
\bea
L_{-e+\delta}&=&K_X(-q_1-\cdots-q_{g-1}-(g-1)p_\infty),
\non
\\
L_{e+\delta}&=&{\cal O}(q_1+\cdots+q_{g-1}-(g-1)p_\infty).
\non
\ena
Therefore 
\bea
(L_{\Delta}\otimes L_{-e})(np_\infty)&\simeq& L_{-e+\delta}\left((g-1+n)p_\infty\right),
\label{ld-lmc}
\\
(L_{\Delta}\otimes L_{e})(np_\infty)&\simeq& L_{e+\delta}\left((g-1+n)p_\infty\right).
\label{ld-lc}
\ena
Let 
\bea
&&
0\leq b_1'<\cdots<b_g'\leq 2g-1,
\non
\\
&&
0\leq b_1^{'\ast}<b_2^{'\ast}<\cdots,
\non
\ena
be gaps and non-gaps of $L_{-e+\delta}$ at $p_\infty$ respectively.

By (\ref{ld-lmc}) we have 
\bea
&&
\dim\left(\text{Ker}\,\pi_U\right)=\sharp\{i\,|\, b_i^{'\ast}-(g-1)\leq 0\},
\qquad
\dim\left(\text{Coker}\,\pi_U\right)=\sharp\{i\,|\, b_i'-(g-1)>0\}.
\non
\ena
Since the number of gaps is $g$ we have 
\bea
&&
\dim\left(\text{Coker}\,\pi_U\right)=g-\sharp\{i\,|\, b_i'-(g-1)\leq 0\}.
\non
\ena
Then 
\bea
&&
\dim\left(\text{Ker}\,\pi_U\right)-\dim\left(\text{Coker}\,\pi_U\right)
\non
\\
&=&\sharp\{i\,|\, b_i^{'\ast}-(g-1)\leq 0\}+\sharp\{i\,|\, b_i'-(g-1)\leq 0\}-g
\non
\\
&=&g-g=0.
\non
\ena
Thus $U\in UGM$. 

By the definition of gaps of $L_{-e+\delta}$ and  
(\ref{ld-lmc}) the sequence $\rho=(\rho(i))_{i<0}$ corresponding to $U$ is given by
\bea
&&
\rho(-i)=-b_i^{' \ast}+g-1.
\label{rho-b-dash}
\ena

In order to prove that $U$ belongs to $UGM^\lambda$ we need to rewrite this in terms of gaps 
of $L_{e+\delta}$.

\begin{lemma}\label{lemma4}
(i)  $(b_1',...,b_g')=(2g-1-b_g^\ast,...,2g-1-b_1^\ast)$.
\vskip2mm
\noindent
(ii)  $(b_1^{'\ast},...,b_g^{'\ast})=(2g-1-b_g,...,2g-1-b_1)$.
\end{lemma}
\vskip2mm
\noindent
{\it Proof.} For simplicity we denote $L_{\pm e+\delta}(kp_\infty)$ by $L_{\pm e+\delta}(k)$ respectively.
By Riemann-Roch theorem, (\ref{ld-lmc}),  (\ref{ld-lc})  we have
\bea
&&
h^0(L_{e+\delta}(g-1+n))-h^0(L_{-e+\delta}(g-1-n))=n.
\label{rel-ML}
\ena
This equation implies that $g-1+n$ is a gap  of $L_{e+\delta}$ if and only if $g-n$ is a non-gap 
of $L_{-e+\delta}$. In fact the condition that $g-1+n$ is a gap of $L_{e+\delta}$ is equivalent to the equation 
\bea
&&
h^0(L_{e+\delta}(g-1+(n-1)))=h^0(L_{e+\delta}(g-1+n)),
\non
\ena
which, by (\ref{rel-ML}), is equivalent to
\bea
&&
h^0(L_{-e+\delta}(g-n))=h^0(L_{-e+\delta}(g-n-1))+1.
\non
\ena
The last equation is equivalent to the condition that $g-n$ is a non-gap of $L_{-e+\delta}$.

Let $n$ be defined by $g-1+n=b_i$. Since $b_i$ is a gap of $L_{e+\delta}$, 
\bea
&&
g-(b_i-g+1)=2g-1-b_i
\non
\ena
 is a non-gap of $L_{-e+\delta}$. Similarly $2g-1-b_i^\ast$ is a gap of $L_{-e+\delta}$ for any $i$. Since 
\bea
&&
0\leq 2g-1-b_i, 2g-1-b_i^\ast\leq 2g-1, \qquad 1\leq i\leq g,
\non
\ena
and these numbers are all distinct, 
they exhaust all gaps and non-gaps of $L_{-e+\delta}$ contained in $\{0,1,...,2g-1\}$.
Thus the lemma is proved.
 $\Box$
\vskip2mm

Notice that $b_i^\ast=i+g-1$ for $i\geq g+1$. Then we have, by (2) of Lemma \ref{lemma4}, 
\bea
\rho&=&(g-1-b_1^{'\ast},...,g-1-b_{g}^{'\ast},-(g+1), -(g+2),...)
\non
\\
&=&(b_g-g,...,b_1-g,-(g+1),-(g+2),...).
\non
\ena
Thus 
\bea
&&
\lambda_\rho=\rho+(1,2,3,...)=(b_g-(g-1),...,b_1,0,0,...)=\lambda,
\non
\ena
which completes the proof of Theorem \ref{U-UGM}.  $\Box$
\vskip2mm

Now we have

\begin{theorem}\label{tau-utau} Let $\tau(t)$ be the solution of the KP-hierarchy given by
 (\ref{theta-sol-KP}) and $U$ the point of UGM given by (\ref{def-U}).
Then 
\bea
&&
C\tau(t)=\tau(t;U)
\label{theta-sol-ntau}
\ena
 for some non-zero constant $C$.
\end{theorem}
\vskip2mm
\noindent
{\it Proof.} 
Notice that the factors of automorphy of $f(p)$ in (\ref{trf-rule-delta-c}) do
 not depend on $t$. Thus if we expand (\ref{part-psi}) in $t={}^t(t_1,t_2,..)$, 
all the coefficients are sections of $I^\ast (\Theta)\otimes L_{-e}$. It implies that if we expand 
$z^{-1}\Psi(t;z)\sqrt{dz}\,\theta(At+e)$ in $t$,  any coefficient is an element of 
$H^0(X, L_{\Delta}\otimes L_{-e}(\ast p_\infty))$.

Let $m_0$ be the order of the zero of $\tau(x,0,...)$ at $x=0$. We expand $x^{m_0}\Psi(x,0,...;z)$ as in (\ref{expansion-psi}).
Then 
\bea
&&
z^{-1}\Psi_i(z)\sqrt{dz}\in H^0(X, L_{\Delta}\otimes L_{-e}(\ast p_\infty)),
\label{BA-expansion}
\ena
for any $i$. Let  
\bea
&&
\Psi_i(z)=\sum_{-\infty<<n<\infty}\Psi_{in}z^n.
\non
\ena
Then 
\bea
&&
\Psi_i(\partial^{-1})e_{-1}=\sum_{-\infty<<n<\infty}\Psi_{in}e_{n-1}.
\label{partial-psi-i}
\ena
On the other hand we have, by the definition (\ref{def-iota}) of the map $\iota$, 
\bea
&&
\iota(z^{-1}\Psi_i(z)\sqrt{dz})=\iota(\sum_{-\infty<<n<\infty}\Psi_{in}z^{n-1}\sqrt{dz})
=\sum_{-\infty<<n<\infty}\Psi_{in}e_{n-1}.
\label{iota-psi-i}
\ena
Let 
\bea
&&
U'=\text{Span}_{\mathbb C}\{\Psi(\partial^{-1})e_{-1}\,|\, i\geq 0\,\}.
\non
\ena
Then $U'$ is the point of $UGM$ corresponding to the solution (\ref{theta-sol-KP}) by Proposition \ref{UGM-psi} and Theorem \ref{UGMpt-tau}.
By (\ref{BA-expansion}),  (\ref{partial-psi-i}) and  (\ref{iota-psi-i}) we have $U'\subset U$.
Since a strict inclusion relation is impossible for two points of UGM we have 
$U'=U$. Thus the theorem is proved. $\Box$

As a corollary of Theorem \ref{tau-utau} we have, by (\ref{exp-tau-lambda}),

\begin{cor}\label{theta-tau-exp}  The $\tau(t)$ given by (\ref{theta-sol-KP})  
has the expansion of the form 
\bea
&&
C\tau(t)=s_{\lambda}(t)+\sum_{\lambda<\mu}\xi_\mu s_\mu(t),
\non
\ena
for some constant $C$.
\end{cor}

%%%%%%%%%%%%%%%%%%%%%%%%%%%%%%%%%%%%%%%%%%%%%%%%%%%%%%%%%%%%%%%%%%%%%%%%%%%%%%%
\section{Series expansion of the theta function}
In this section we assume that a data $(X,\{\alpha_i,\beta_i\}, p_\infty, z,e)$ is given, where 
 notation is as before.

\subsection{Taylor expansion}
In order to study the series expansion of the theta function we need to change the variables 
to appropriate ones. 

Let $\{du_{w_i}\}$ be a basis of holomorphic one forms such that $du_{w_i}$ has the 
following expansion at $p_\infty$:
\bea
&&
du_{w_i}=z^{w_i-1}(1+O(z)) dz.
\label{du-exp}
\ena
The existence of such a basis is easily proved using Riemann-Roch theorem (c.f. \cite{KS}).
However a basis which satisfies the condition (\ref{du-exp}) is not unique. 
For the moment we do not know what is the best choice. So we take 
any one of them.

By integrating this basis over the canonical homology basis we define the period matrices $\omega_1$,
$\omega_2$ by
\bea
&&
2\omega_1=\left(\int_{\alpha_j}du_i\right),\qquad 2\omega_2=\left(\int_{\beta_j}du_i\right).
\non
\ena
The normalized period matrix $\Omega$ is given by $\Omega=\omega_1^{-1}\omega_2$.

We label the $g$ variables of the theta function by gaps at $p_\infty$ as 
\bea
&&
u={}^t(u_{w_1},...,u_{w_g}),
\non
\ena
and consider the function of the form 
\bea
&&
\theta\bigl((2\omega_1)^{-1}u+e\,|\,\Omega\,\bigr).
\non
\ena
We set $\partial_{i}=\partial/\partial u_i$ and $\partial_I=\partial_{i_1}\cdots\partial_{i_r}$ for 
$I=(i_1,...,i_r)$. If we differentiate theta function $i_j$ should be considered in $\{w_1,...,w_g\}$.
We assign weight $i$ to $u_i$.
Then the initial term, with respect to weight, of the Taylor series expansion of this function is given by the following theorem.

\begin{theorem}\label{expansion-theta} Suppose that a data 
$(X, \{\alpha_i,\beta_i\}, p_\infty, z ,\{du_{w_i}\}, e)$ is 
given. Let $\lambda$ be the partition associated with it. 
Then we have
\bea
&&
C\theta\bigl((2\omega_1)^{-1}u+e\,|\,\Omega\,\bigr)=s_\lambda(t)|_{t_{w_i}=u_{w_i}}+\text{higher weight terms}.
\non
\ena
where $C=c_0\partial_{A_0}\theta(e|\Omega)$ and $c_0=\pm1$ is given by (\ref{signature}).
\end{theorem}
\vskip2mm
\noindent
{\it Proof.} Let us expand $du_{w_i}$ at $p_\infty$ in $z$ as 
\bea
&&
du_{w_i}=\sum_{j=1}^\infty b_{ij} z^{j-1} dz,
\non
\ena
and define the matrix $B=(b_{ij})$. By (\ref{du-exp}) $B$ has the following triangular 
structure:
\bea
&&
b_{ij}=\left\{
\begin{array}{cr}
0& \quad\text{if $j<w_i\,$}\\
1& \quad\,\, \text{ if $j=w_i$.}\\
\end{array}\right.
\label{triangular-B}
\ena

\begin{lemma}\label{AB-relation} 
We have $B=2\omega_1A$.
\end{lemma}
\vskip2mm
\noindent
{\it Proof.} Write 
\bea
&&
du_{w_i}=\sum_{j=1}^g c_{ij} dv_j.
\label{du-dv-rel}
\ena
Integrating this equation over $\alpha_j$ we get $(2\omega_1)_{ij}=c_{ij}$.
Then the lemma follows by comparing the expansion 
coefficients of (\ref{du-dv-rel}). $\Box$
\vskip2mm

By the lemma the tau function (\ref{theta-sol-KP}) can be written as 
\bea
&&
\tau(t)=e^{\frac{1}{2}q(t)}\theta((2\omega_1)^{-1}Bt+e\,|\,\Omega).
\label{Bform-tau}
\ena
Let $u={}^t(u_{w_1},...,u_{w_g})=Bt$ and set $t_j=0$ for all $j$ except $t_{w_i}$, $1\leq i\leq g$.
Then we can write  
\bea
&&
u=\tilde{B}\tilde{t},\qquad \tilde{t}={}^t(t_{w_1},...,t_{w_g}),
\non
\ena
where the $g\times g$ matrix $\tilde{B}=(\tilde{b}_{ij})$ is the upper triangular matrix whose
diagonal entries are all equal to $1$ due to (\ref{triangular-B}). Thus $\tilde{B}$ is invertible 
and $\tilde{B}^{-1}$ has a similar form to $\tilde{B}$.
Then 
\bea
&&
\theta((2\omega_1)^{-1}u+e\,|\,\Omega)=e^{-\frac{1}{2}q(\tilde{u})}\tau(\tilde{u}),
\label{theta-tau}
\\
&&
\tilde{u}={}^t(\tilde{u}_{w_{1}},...,\tilde{u}_{w_{g}})=\tilde{B}^{-1}u
\non
\ena
where $\tilde{u}$ in $\tau(t)$and $q(t)$ should be understood that $\tilde{u}_{w_{i}}$ sits on the $w_i$-th 
component of $t={}^t(t_1,t_2,...)$ and other components of $t$ are zero.
Since
 $\tilde{u}_{w_i}=
u_{w_i}$ modulo higher weight terms, the theorem follows from Corollary 
\ref{theta-tau-exp} and Theorem \ref{th-2} (iii). $\Box$

%%%%%%%%%%%%%%%%%%%%%%%%%%%%%%%%%%%%%%%%%%%%%%%%%%%%%%%%%%%%%%%%%%%%%%%%%%%%%%%%%%%%%%%
\subsection{Duality}
There is a relation of the expansions of the theta function at $e$ and at $-e$.

\begin{theorem}\label{expansion-theta-2} Under the same assumption as in Theorem \ref{expansion-theta}
we have 
\bea
&&
(-1)^{|\lambda|}C\theta\bigl((2\omega_1)^{-1}u-e\,|\,\Omega\,\bigr)=s_{{}^t\lambda}(t)|_{t_{w_i}=u_{w_i}}+\text{higher weight terms},
\non
\ena
where ${}^t\lambda$ is the conjugate partition of $\lambda$ and $C$ is the same as in Theorem \ref{expansion-theta}.
\end{theorem}
\vskip2mm
\noindent
{\it Proof.} Since $\theta(Z\,|\,\Omega)$ is an even function of $Z$ 
\bea
&&
\theta\bigl((2\omega_1)^{-1}u-e\,|\,\Omega\,\bigr)=\theta\bigl((2\omega_1)^{-1}(-u)+e\,|\,\Omega\,\bigr).
\non
\ena
Then the theorem follows from the known relation for the Schur functions \cite{Mac}
\bea
&&
s_\mu(-t)=(-1)^{|\mu|}s_{{}^t\mu}(t),
\non
\ena
and Theorem \ref{expansion-theta}.
$\Box$

\begin{cor}\label{cor2} Let $\lambda$ be the partition associated with $(X, \{\alpha_i,\beta_i\}, p_\infty, e)$.
Then the partition associated with $(X, \{\alpha_i,\beta_i\}, p_\infty, -e)$ is ${}^t\lambda$.
\end{cor}
\vskip2mm
\noindent
{\it Proof.} Since $s_\mu(t)=s_\nu(t)$ implies $\mu=\nu$ for two partitions $\mu$ and $\nu$,
the assertion follows from Theorem \ref{expansion-theta} and Theorem \ref{expansion-theta-2}.
$\Box$

%%%%%%%%%%%%%%%%%%%%%%%%%%%%%%%%%%%%%%%%%%%%%%%%%%%%%%%%%%%%%%%%%%%%%%%%%%%%%%%%%%%%%%%%%%
\subsection{Examples}
We give examples of the partition $\lambda$ given by (\ref{lambda}) which, by Theorem \ref{expansion-theta}, 
determines the initial term of the Taylor expansion of the theta function.
\vskip3mm
\noindent
{\bf Example 1} Let $X$ be any compact Riemann surface of genus $g$ , $p_\infty$ a non-Weierstrass 
point of $X$ and $e=-\delta$. In this case $L_{e+\delta}\simeq {\cal O}$ and we have
\bea
&&
(b_1,...,b_g)=(w_1,...,w_g)=(1,2,...,g).
\non
\ena
Thus $\lambda=(1^g)$. In this case $s_{\lambda}(t)=(-1)^g p_g (-t)$. For example, for 
 $1\leq g\leq 4$ they are given by
\bea
&&
s_{(1)}(t)=t_1, \qquad s_{(1,1)}(t)=-t_2+\frac{t_1^2}{2}, \qquad 
s_{(1,1,1)}(t)=t_3-t_1t_2+\frac{t_1^3}{3!},
\non
\\
&&
s_{(1,1,1,1)}(t)=-t_4+t_1t_3+\frac{t_2^2}{2}-\frac{t_1^2t_2}{2}+\frac{t_1^4}{4!}.
\non
\ena
The $(3,4,5)$ curve studied in \cite{KM} is included in this Example.
\vskip5mm
\noindent
{\bf Example 2} Let $X$ and $p_\infty$  be the same as in Example 1. Take $e=\delta$. Then, by duality,
$\lambda=(g)$. In this case $s_{\lambda}(t)=p_g(t)$.
\vskip5mm
\noindent
{\bf Example 3} Let $X$ be an $(n,s)$ curve \cite{BEL2} or a telescopic curve of genus $g$ 
\cite{Miu}\cite{A}\cite{AN}. Take $p_\infty=\infty$ and $e=-\delta$. Then 
\bea
&&
\lambda=(w_g,...,w_1)-(g-1,...,1,0).
\label{usual}
\ena
In this case $\delta=-\delta$ in $J(X)$ and $\lambda$ satisfies ${}^t\lambda=\lambda$ by Corollary \ref{cor2}. These are cases which were studied in most of the literatures.
\vskip5mm
\noindent
{\bf Example 4} In \cite{KMP} sigma functions of $(3,7,8)$ $(g=4)$ and $(6,13,14,15,16)$ $(g=12)$ 
curves are studied. If we take $e=-\delta$, then (\ref{usual}) holds.
For $(3,7,8)$-curve, gaps are (1,2,4,5) and $\lambda=(2,2,1,1)$. For $(6,13,14,15,16)$-curve, gaps are 
(1,2,3,4,5,7,8,9,10,11,17,23) and $\lambda=(12,7,2,2,2,2,2,1,1,1,1,1)$. In the latter case the maximal 
gap $23$ is equal to $2g-1$. It implies $\delta=-\delta$ in $J(X)$ and $\lambda={}^t\lambda$.

If we take $e=\delta$, then we have $\lambda=(4,2)$ for $(3,7,8)$-curve by duality.
\vskip2mm

%%%%%%%%%%%%%%%%%%%%%%%%%%%%%%%%%%%%%%%%%%%%%%%%%%%%%%%%%%%%%%%%%%%%%%%%%%%%%%%%%%%%%%%%%%
\subsection{The case of hyperelliptic curves}
In the case of hyperelliptic curves it is possible to determine the partition $\lambda$ corresponding to any point on the theta divisor explicitly.

Let $X$ be the hyperelliptic curve of genus $g$ given by 
\bea
&&
y^2=f(x), \qquad f(x)=\prod_{i=1}^{2g+1}(x-e_i),
\non
\ena
where $e_i\neq e_i$ for any $i\neq j$. We take a canonical homology basis $\{\alpha_i,\beta_i\}$ and 
the base point $p_\infty=\infty$. 
Let $\phi$ be the hyperelliptic involution, $\phi(x,y)=(x,-y)$. 
It is known  that the points on the theta divisor with the multiplicity 
$m_0\geq 1$ are given by
\bea
&&
q_1+\cdots+q_{g+1-2m_0}+(2m_0-2)\infty -\Delta, 
\label{mult-m0}
\ena
where $q_i's$ are points on $X$ satisfying $q_i\neq \phi(q_j)$ for any $i\neq j$ \cite{F}. We denote this set of points by $\Theta_{m_0}$.  

We set
\bea
\Theta_{m_0,o}&=&\{q_1+\cdots+q_{g+1-2m_0}+(2m_0-2)\infty-\Delta\in \Theta_{m_0}\,\vert\,
 q_i\neq \infty\,\, \forall i\,\},
\non
\\
\Theta_{m_0,e}&=&\{q_1+\cdots+q_{g-2m_0}+(2m_0-1)\infty-\Delta\in \Theta_{m_0}\,\vert\,
q_i\neq \infty\,\, \forall i \,\}.
\non
\ena
Then

\begin{prop}\label{part-hyperelliptic} 
The partition $\lambda$ associated with $(X, \{\alpha_i,\beta_i\}, \infty, e)$ is given as follows.
\vskip2mm
\noindent
(i) $\lambda=(2m_0-1,2m_0-2,...,1)$ \,for \, $e\in \Theta_{m_0,o}$.
\vskip2mm
\noindent
(ii) $\lambda=(2m_0,2m_0-1,...,1)$ \,for \,$e\in \Theta_{m_0,e}$.
\end{prop}

This proposition follows from 

\begin{lemma} The gaps of $L_{e+\delta}$ are given as follows.
\vskip4mm
\noindent
(i) $\{\,0,1,...,g-2m_0, g-2m_0+2i\,\, (1\leq i\leq 2m_0-1)\}$ \,for \, $e\in \Theta_{m_0,o}$,
\vskip2mm
\noindent
(ii) $\{\,0,1,...,g-1-2m_0, g-1-2m_0+2i\,\, (1\leq i\leq 2m_0)\}$ \,for \, $e\in \Theta_{m_0,e}$.
\end{lemma}
\vskip2mm
\noindent
{\it Proof.} Let us prove (i).  The proof of (ii) is similar. 
We have 
\bea
&&
e+\delta=q_1+\cdots+q_{g+1-2m_0}-(g+1-2m_0)\infty.
\non
\ena
Since 
\bea
&&
L_{e+\delta}(n\infty)\simeq {\cal O}_X\bigl(q_1+\cdots+q_{g+1-2m_0}+(n-(g+1-2m_0))\infty\bigr),
\non
\ena
 $H^0(X,L_{e+\delta}(n\infty))$ is identified with the space of 
meromorphic functions on $X$ with at most a simple pole at $q_i$ $(1\leq i\leq g+1-2m_0)$ and a pole of order 
at most $n-(g+1-2m_0)$ at $\infty$ if $n-(g+1-2m_0)\geq 0$ or with at most a simple pole at $q_i$ $(1\leq i\leq g+1-2m_0)$ and a zero of order at least $-n+(g+1-2m_0)$ at $\infty$ if $n-(g+1-2m_0)< 0$.

Let $q_i=(x_i,y_i)$.
Since a meromorphic function on $X$ which has a pole only at $\infty$ is a polynomial of $x$ and $y$, 
any element of $H^0(X,L_{e+\delta}(n\infty))$ can be written in the form 
\bea
&&
F(x,y)=\frac{A(x)+B(x)y}{(x-x_1)\cdots(x-x_{g+1-2m_0})},
\label{section-F}
\ena
$A(x)$ and $B(x)$ are some polynomials of $x$. Since $F$ has at most a simple pole  at $q_i=(x_i,y_i)$, 
$A(x)$ and $B(x)$ should satisfy
\bea
&&
A(x_i)-B(x_i)y_i=0, \quad 1\leq i\leq g+1-2m_0.
\label{cond-simple-pole}
\ena

First consider the case $n<g+1-2m_0$. In this case $F$ must have a zero at $\infty$ of order at 
least $g+1-2m_0-n$. Let us estimate the order of zero or pole at $\infty$ of $F$ given by (\ref{section-F}).

If $B(x)=0$, then by (\ref{cond-simple-pole}), $A(x)$ must be divided by $\prod_{i=1}^{g+1-2m_0}(x-x_i)$
and $F$ becomes a polynomial of $x$. Thus $F$ can not have a zero at $\infty$.

Suppose that $B(x)\neq 0$. Let $k$ and $l$ be degrees of $A(x)$ and $B(x)$ respectively. 
Since $x$ and $y$ has a pole of order $2$ and $2g+1$ at $\infty$, 
the order of a zero of $F$ at $\infty$, denoted by $\text{ord}(F)$, is given by
\bea
&&
\text{ord}(F)=
2(g+1-2m_0)-\max\{2k,2l+2g+1\}.
\non
\ena
Notice that $\max\{2k,2l+2g+1\}\geq 2g+1$. Then 
\bea
&&
\text{ord}(F)
\leq 
2(g+1-2m_0)-(2g+1)
=
1-4m_0<0.
\non
\ena
Thus $F$ can not have a zero at $\infty$ in this case too. 
Consequently $0,1,...,g-2m_0$ are gaps of $L_{e+\delta}$ at $\infty$.

If $n=g+1-2m_0$, the constant function $1$ can be considered as a global section of $L_{e+\delta}(n\infty)$. Thus 
$n$ is a non-gap.

Suppose that $n>g+1-2m_0$. In this case $x^{2i}$, $i\geq 1$ gives a global section of  $L_{e+\delta}(n\infty)$ 
with $n=2i+(g+1-2m_0)$. Let us prove that $n=2i-1+(g+1-2m_0)$, $1\leq i\leq 2m_0-1$ is a 
gap. The order of a pole of $F$ at $\infty$, which is denoted by $-\text{ord}(F)$, is given by
\bea
&&
-\text{ord}(F)=\max\{2k,2l+2g+1\}-2(g+1-2m_0)\geq 2l-1+4m_0.
\non
\ena
Therefore, if $n=2i-1+(g+1-2m_0)$ is a non gap,  it must satisfy
\bea
&&
2i-1\geq 2l-1+4m_0,
\non
\ena
which implies $i\geq 2m_0+l\geq 2m_0$. Thus  $n=2i-1+(g+1-2m_0)$, $1\leq i\leq 2m_0-1$ is a 
gap. 

The number of gaps we obtained so far is $(g-2m_0+1)+(2m_0-1)=g$. Thus those exhaust all gaps. 
Thereby the proof of the proposition is completed. $\Box$

\vskip3mm
\noindent
{\bf Remark} If $g$ is odd and $m_0=\frac{g+1}{2}$ then $e=-\delta\in \Theta_{m_0,o}$. In this case
$\lambda=(g,g-1,...,1)$. If $g$ is even and $m_0=\frac{g}{2}$, then $e=-\delta\in \Theta_{m_0,e}$ and 
$\lambda=(g,g-1,...,1)$. These results are the same as those given in Example 3 for $(2,2g+1)$ curves 
as it should be.

%%%%%%%%%%%%%%%%%%%%%%%%%%%%%%%%%%%%%%%%%%%%%%%%%%%%%%%%%%%%%%%%%%%%%%%%%%%%%%%%%%%%%%%%
\subsection{Expansion on Abel-Jacobi images}

 We denote by $l(\mu)$ the length of a partition $\mu$.

\begin{prop}\label{prop4} Let $0\leq k\leq l(\lambda)$. Then, for any $I=(i_1,...,i_m)$, $m\geq 1$, satisfying 
$\sum_{j=1}^m i_j<N_{\lambda,k}$ we have 
\bea
&&
\partial_I\theta(\sum_{i=1}^k (p_i-p_\infty)+e\,|\,\Omega)=0,
\non
\ena
for any $p_i\in X$, $1\leq i\leq k$.
\end{prop}
\vskip2mm
\noindent
{\it Proof.} Differentiate (\ref{theta-tau}) by $\partial_I$ and use Proposition 4 in \cite{NY}, Corollary \ref{theta-tau-exp}, to get the result.
 $\Box$

\begin{theorem}\label{theta-AJimage} Suppose that $m_k>0$.  Let $c_k$ be given by (\ref{signature}) and $z_j=z(p_j)$ for $p_j\in X$. Then
\vskip2mm
\noindent
(i) $\displaystyle{C\partial_{A_k}
\theta(\sum_{j=1}^k (p_j-p_\infty)+e\,|\,\Omega)=c_ks_{(\lambda_1,...,\lambda_k)}(\sum_{j=1}^k[z_j] )+
\text{higher weight terms},}$

where $C$ is the constant in Theorem \ref{expansion-theta} and $c_k=\pm1$ is given by (\ref{signature}).
\vskip2mm
\noindent
(ii) $\displaystyle{
\partial_{A_k}
\theta(\sum_{j=1}^k (p_j-p_\infty)+e\,|\,\Omega)}$

\hskip35mm
$\displaystyle{
=\frac{c_k}{c_{k-1}}\partial_{A_{k-1}}
\theta(\sum_{j=1}^{k-1} (p_j-p_\infty)+e\,|\,\Omega)z_k^{\lambda_k}+O(z_k^{\lambda_k+1}),}$

where $z_j=z(p_j)$.
\end{theorem}
\vskip2mm
\noindent
{\it Proof.} The theorem easilly follows from Corollary \ref{theta-tau-exp}, Theorem \ref{tau-AYimage} and (\ref{Bform-tau}) 
as in the proof of Corollary 4 in \cite{NY}. $\Box$

%%%%%%%%%%%%%%%%%%%%%%%%%%%%%%%%%%%%%%%%%%%%%%%%%%%%%%%%%%%%%%%%%%%%%%%%%%%%%%%%%%%%%%%%%%
\subsection{Refined Riemann's singularity theorem}

\begin{cor}\label{refined-RST} We assume the same conditions as in Theorem \ref{expansion-theta}. Then
\vskip2mm
\noindent
(i) For any $I=(i_1,...,i_m)$, $m\geq 1$, satisfying $\sum_{j=1}^m i_j<|\lambda|$ we have 
\bea
&&
\partial_I\theta(e\,|\,\Omega)=0.
\non
\ena
\vskip2mm
\noindent
(ii) If $m<m_0$ we have 
\bea
&&
\partial_I\theta(e\,|\,\Omega)=0,
\non
\ena
 for any $I=(i_1,...,i_m)$.
\vskip2mm
\noindent
(iii) $\displaystyle{\partial_{A_0}\theta(e\,|\,\Omega)}\neq0$.
\end{cor}
\vskip2mm
\noindent
{\it Proof.} (i) is the special case $k=0$ of Proposition \ref{prop4}.
(ii) follows from Theorem \ref{Rvanishing-tau} and (\ref{theta-tau}). (iii) is the special case 
$k=0$ of Theorem \ref{theta-AJimage} (i). $\Box$
\vskip2mm

Notice that Riemann's singularity theorem is a part of this corollary. 
Moreover the assertion  (iii) gives one non-vanishing 
derivative with degree $m_0$ explicitly and (i) gives a new vanishing property of the theta function 
which does not follow from Riemann's singularity theorem.  In this sense Corollary \ref{refined-RST}
is an extension and a refinement of Riemann's singularity theorem. By Proposition \ref{minimal-wt-deg}, we can say more.

\begin{cor} In the series expansion of the theta function $C\theta((2\omega_1)^{-1}u+e\,|\Omega)$ at $u=0$ 
the terms with the minimal weight and the minimal degree are given by 
$L_\lambda(t)|_{t_{w_i}=u_{w_i}}$, where $C$ is the constant in Theorem \ref{expansion-theta}.
\end{cor}

%%%%%%%%%%%%%%%%%%%%%%%%%%%%%%%%%%%%%%%%%%%%%%%%%%%%%%%%%%%%%%%%%%%%%%%%%%%%%%%%%%%
\section{Sigma function}
Sigma functions of an arbitrary Riemann surface have been introduced by Korotkin and Shramchenko\cite{KS}. In this section we first review their construction in a more 
unified frame work. Next we prove the modular invariance of  sigma functions
with characteristics by specifying a suitable normalization constant which is obtained as a 
result of  Corollary \ref{refined-RST} and is apparently different from that in \cite{KS}. 

In this section we assume that a data $(X, \{\alpha_i,\beta_i\}, p_\infty, z, e,\{du_{w_i}\})$ is given. 

\subsection{Bilinear meromorphic differential}

The key ingredients in constructing sigma function are certain 
bilinear meromorphic differentials. So we first study  general properties of 
such bilinear differentials.

Let $\widehat{\omega}(p_1,p_2)$ be a  symmetric bilinear meromorphic differential on $X\times X$ 
such that it is holomorphic outside the diagonal $\{(p,p)|\,p\in X\}$ where it has a double pole 
and at any $p_0\in X$ it has the expansion of the form
\bea
&&
\widehat{\omega}(p_1,p_2)=\left(\frac{1}{(w_1-w_2)^2}+\text{holomorphic in $w_1,w_2$}\right)dw_1dw_2,
\non
\ena
where $w$ is a local coordinate at $p_0$ and $w_i=w(p_i)$. 

The fundamental normalized differential of the second kind $\omega(p_1,p_2)$ is an example (see (\ref{bilinear-omega})).
In general $\widehat{\omega}(p_1,p_2)$ has the following structure.

\begin{prop}\label{prop-decomp}
For any $\widehat{\omega}(p_1,p_2)$ there exist $\Omega(p_1,p_2)$ and $\{d\tilde{r}_i\}_{i=1}^\infty$ such that
\bea
&&
\widehat{\omega}(p_1,p_2)=d_{p_2}\widehat{\Omega}(p_1,p_2)+\sum_{i=1}^g du_{w_i}(p_1)d\tilde{r}_i(p_2).
\label{decomp-homega}
\ena
Here $d\tilde{r}_i(p)$ is a locally exact meromorphic one form on $X$ 
which has a pole only at $p_\infty$ and $\widehat{\Omega}(p_1,p_2)$ is a meromorphic one form on $X\times X$  which satisfies the following conditions.
\vskip2mm
\noindent
(i) It is a meromorphic one form in $p_1$ for a fixed $p_2$.
\vskip2mm
\noindent
(ii) It is a meromorphic function in $p_2$ for a fixed $p_1$.
\vskip2mm
\noindent
(iii) It is holomorphic except $\{(p,p)|\,p\in X\}\cup \{(p_\infty,p)|\,p\in X\}\cup \{(p,p_\infty)|\,p\in X\}$.
\vskip2mm
\noindent
(iv) It has a simple pole at the diagonal $\{(p,p)|\,p\in X\}$.
\end{prop}
\vskip2mm
\noindent
{\it Proof.} By Lemma 7 of \cite{N1} $\widehat{\omega}$ can be written in terms of $\omega$ as 
\bea
&&
\widehat{\omega}(p_1,p_2)=\omega(p_1,p_2)+\sum_{i,j=1}^g c_{ij} du_{w_i}(p_1)du_{w_j}(p_2),
\non
\ena
for some constants $c_{ij}$ satisfying $c_{ij}=c_{ji}$. Therefore it is sufficient to prove (\ref{decomp-homega})  for $\omega$.

Let us consider the exact sequence of sheaves
\bea
&&
0\longrightarrow {\mathbb C} \longrightarrow {\cal O}(\ast p_\infty)\longrightarrow d{\cal O}(\ast p_\infty)\longrightarrow 0,
\non
\ena
where $d:{\cal O}(\ast p_\infty)\rightarrow K_X(\ast p_\infty)$ is the exterior differentiation.
Taking cohomologies we have
\bea
&&
H^1(X,{\mathbb C})\simeq H^0(X,d{\cal O}(\ast p_\infty))/dH^0(X,{\cal O}(\ast p_\infty)).
\label{homology-2ndkinddiff}
\ena

We remark that $ H^0(X,d{\cal O}(\ast p_\infty))$ is the space of locally exact meromorphic one forms 
with a pole only at $p_\infty$. Thus this space is generated, as a vector space, 
 by holomorphic one forms and the 
differentials of the second kind with a pole only at $p_\infty$. We realize $H^1(X,{\mathbb C})$ as 
in the right hand side of (\ref{homology-2ndkinddiff}).

Let $\{du^{(1)}_i, du^{(2)}_i\}$ be the basis of $H^1(X,{\mathbb C})$ which is 
 dual to  $\{\alpha_i,\beta_i\}$. Namely they satisfy
\bea
&&
\int_{\alpha_j}du^{(1)}_i=\delta_{ij}, \qquad \int_{\alpha_j}du^{(2)}_i=0,
\non
\\
&&
\int_{\beta_j}du^{(1)}_i=0, \qquad \int_{\beta_j}du^{(2)}_i=\delta_{ij}.
\label{dual-basis}
\ena
We set
\bea
&&
\widehat{\Omega}(p_1,p_2)=d_{p_1}\log\frac{E(p_1,p_2)}{E(p_1,p_\infty)}-2\pi i\sum_{k=1}^g dv_k(p_1)\int^{p_2}
du^{(2)}_k.
\label{large-Omega}
\ena
We show that this $\widehat{\Omega}(p_1,p_2)$ satisfies properties (i) - (iv).

Since the first term in the right hand side is invariant when $p_1$ goes round $\alpha_i$ or $\beta_i$ cycles due to the 
transformation rule of the theta function, (i) is obvious.  Since $E(p_1,p_2)$ 
has a unique simple zero at
 $p_1=p_2$, (iii) and (iv) are also obvious. Let us prove (ii). 
We have to show that $\widehat{\Omega}(p_1,p_2)$ is 
invariant when $p_2$ goes round $\alpha_i$ or $\beta_i$.

  If $p_2$ goes round $\alpha_j$ then $\widehat{\Omega}(p_1,p_2)$ is invariant 
due to the transformation rule of the theta function and (\ref{dual-basis}). Next consider the case 
where  $p_2$ goes round $\beta_j$. Then the integral of the normalized holomorphic differential changes as 
\bea
&&
\int_{p_1}^{p_2} dv \mapsto \int_{p_1}^{p_2} dv+\Omega {\bf e}_j,
\non
\ena
where ${\bf e}_j={}^t(0,...,1,...,0)$ is the $j$-th unit vector and $\Omega$ 
is the normalized period matrix. 
Since
\bea
&&
\theta[\varepsilon_o](\int_{p_1}^{p_2} dv+\Omega {\bf e}_j)
\non
\\
&&=
e^{-2\pi i(\varepsilon_o'')_j-\pi i\Omega_{jj}-2\pi i\int_{p_1}^{p_2} dv_j}
\theta[\varepsilon_o](\int_{p_1}^{p_2} dv),
\non
\ena
we have 
\bea
&&
d_{p_1}\log\frac{E(p_1,p_2)}{E(p_1,p_\infty)}\mapsto 
d_{p_1}\log\frac{E(p_1,p_2)}{E(p_1,p_\infty)}+2\pi i dv_j(p_1).
\label{change-1term}
\ena
On the other hand we have
\bea
&&
\int^{p_2}du^{(2)}_k \mapsto \int^{p_2}du^{(2)}_k+\delta_{jk}.
\non
\ena
Thus 
\bea
&&
-2\pi i\sum_{k=1}^g dv_k(p_1)\int^{p_2}
du^{(2)}_k
\mapsto
-2\pi i\sum_{k=1}^g dv_k(p_1)\int^{p_2}
du^{(2)}_k
-2\pi i dv_j(p_1).
\label{change-2term}
\ena
By (\ref{change-1term}) and (\ref{change-2term}) $\widehat{\Omega}(p_1,p_2)$ is invariant when $p_2$ goes 
round $\beta_j$.
 Thus the property (ii) is proved.

Next let us prove (\ref{decomp-homega}) by defining $d\tilde{r}_i$ explicitly.
 Taking $d_{p_2}$ of  (\ref{large-Omega}) we get 
\bea
&&
d_{p_2}\widehat{\Omega}(p_1,p_2)=\omega(p_1,p_2)-2\pi i\sum_{j=1}^g dv_j(p_1)du^{(2)}_j(p_2).
\label{dp2}
\ena
Using $dv_i=\sum_{j=1}^g c_{ij}du_{w_j}$, $c_{ij}=( (2\omega_1)^{-1})_{ij}$, we have 
\bea
&&
2\pi i\sum_{j=1}^g dv_j(p_1)du^{(2)}_j(p_2)=\sum_{j=1}^gdu_{w_j}(p_1)d\tilde{r}_j(p_2),
\qquad
 d\tilde{r}_j:=2\pi i\sum_{k=1}^g c_{kj}du^{(2)}_k.
\non
\ena
Since $d\tilde{r}_j$ is a locally exact meromorphic one form with a pole only at 
$p_\infty$ by construction, the relation  (\ref{decomp-homega}) is proved. $\Box$

\begin{cor} $\{d u_{w_i}, d\tilde{r}_i\}$ is a symplectic basis of $H^1(X,{\mathbb C})$ 
with respect to the intersection form $\bullet$:
\bea
&&
d u_{w_i}\bullet d u_{w_j}=d\tilde{r}_i\bullet d\tilde{r}_j=0, \qquad 
d u_{w_i}\bullet d\tilde{r}_j=\delta_{ij}.
\label{symplectic}
\ena
\end{cor}
\vskip2mm
\noindent
{\it Proof.} 
The computation of the intersection is the same as that in Proposition 3 in \cite{N1} using
 Proposition \ref{prop-decomp}. $\Box$

Let us define the period matrices $\eta_i$, $i=1,2$ by
\bea
&&
-2\eta_1=\left(\int_{\alpha_j}d\tilde{r}_i\right),
\qquad
-2\eta_2=\left(\int_{\beta_j}d\tilde{r}_i\right),
\non
\ena
and set 
\bea
&&
M=\left(\begin{array}{cc}\omega_1&\omega_2\\ \eta_1&\eta_2\end{array}\right).
\non
\ena
As a consequence of (\ref{symplectic}) $M$ satisfies the Riemann bilinear relation 
of the following form \cite{N1}
\bea
{}^tM
\left(\begin{array}{cc}0&I_g\\ -I_g&0\end{array}\right)
M=-\frac{\pi i}{2}\left(\begin{array}{cc}0&I_g\\ -I_g&0\end{array}\right).
\non
\ena
It guarantees the consistency of the transformation rule of sigma functions \cite{BEL1}.

The following corollary shows that we can determine $\eta_1, \eta_2$ directly from 
$\widehat{\omega}(p_1,p_2)$. 

\begin{cor} The following relations are valid:
\bea
&&
\int_{\alpha_j}\widehat{\omega}(p_1,p_2) =
\sum_{i=1}^g du_{w_i}(p_1)(-2\eta_{1,ij}),
\label{eta1}
\\
&&
\int_{\beta_j}\widehat{\omega}(p_1,p_2) =
\sum_{i=1}^g du_{w_i}(p_1)(-2\eta_{2,ij}),
\label{eta2}
\ena
where the integrals are taken with respect to the variable $p_2$.
\end{cor}
\vskip5mm

\subsection{Klein form}
In defining sigma functions we need a meromorphic bilinear differential $\widehat{\omega}(p_1,p_2)$ which is independent 
of the choice of canonical homology basis. The Klein differential \cite{K,F} is one of such differentials.
Let us recall its definition. 

Let ${\cal S}$ be the set of half characteristics $\varepsilon={}^t(\varepsilon',\varepsilon'')$, 
$\varepsilon',\varepsilon''\in(\frac{1}{2} {\mathbb Z}^g)/{\mathbb Z}^g$ such that $\theta[\varepsilon](0|\Omega)$ 
does not vanish and $N({\cal S})$ denote the number of elements of ${\cal S}$. Then the Klein form $\omega_K(p_1,p_2)$ 
is defined by
\bea
&&
\omega_K(p_1,p_2)=\omega(p_1,p_2)+\frac{1}{N({\cal S})}\sum_{i,j=1}^g dv_i(p_1)dv_j(p_2)
\frac{\partial^2}{\partial z_i\partial z_j}\log {\cal F}(z)|_{z=0},
\non
\\
&&
 {\cal F}(z)=\prod_{\varepsilon\in {\cal S}}\theta[\varepsilon](z|\Omega).
\non
\ena
Using the transformation formula of the theta function under the action of the symplectic group, 
it can be proved that $\omega_K$ is independent of the choice of canonical homology basis  \cite{F,KS,K}.

We can easily compute periods of Klein form $\omega_K(p_1,p_2)$ using
\bea
&&
\int_{\alpha_j} \omega(p_1,p_2)=0,
\hskip10mm
\int_{\beta_j} \omega(p_1,p_2)=2\pi i dv_j(p_1),
\non
\ena
where integrals are with respect to the variable $p_2$. 
The result is given by (\ref{eta1}) and (\ref{eta2}) with 
\bea
&&
\eta_1={}^t\omega_1^{-1}\Lambda,
\hskip10mm
\eta_2=-\frac{\pi}{2}{}^t\omega_1^{-1}+{}^t\omega_1^{-1}\Lambda \Omega,
\non
\ena
where $\Lambda=(\Lambda_{ij})$ is given by
\bea
&&
\Lambda_{ij}=-\frac{1}{4N(S)}\frac{\partial^2}{\partial z_i\partial z_j}\log {\cal F}(z)|_{z=0}.
\non
\ena
These $\eta_i$'s  are nothing but the ones given in \cite{KS}.

%%%%%%%%%%%%%%%%%%%%%%%%%%%%%%%%%%%%%%%%%%%%%%%%%%%%%%%%%%%%%%%%%%%%%%%%%%%%%%%%%%%%%%%%%%%
\subsection{Definition of sigma function}
Take any $\widehat{\omega}(p_1,p_2)$ of the form 
\bea
&&
\widehat{\omega}(p_1,p_2)=\omega_K(p_1,p_2)+\sum_{i,j=1}^g c_{ij} du_{w_i}(p_1)du_{w_j}(p_2),
\label{sigma-homega}
\ena
where $c_{ij}=c_{ji}$ is a constant independent of the choice of canonical homology basis.
Define the period matrices $\eta_i$ by (\ref{eta1}) and (\ref{eta2}).
Let $\varepsilon={}^t(\varepsilon',\varepsilon'')$, $\varepsilon',\varepsilon''\in {\mathbb R}^g$ be the characteristics of $e$:
\bea
&&
e=q_1+\cdots+q_{g-1}-\Delta=\Omega \varepsilon'+\varepsilon'',
\non
\ena
and $u={}^t(u_{w_1},...,u_{w_g})$. Notice that if we change the choice of canonical homology basis 
$\{\alpha_i,\beta_i\}$, then $\Delta$, $\Omega$ and the Abel-Jacobi map are changed. Consequently  $\varepsilon$ also depends on the choice of canonical homology basis.

By Corollary \ref{refined-RST} we know 
\bea
&&
\partial_{A_0}\theta[\varepsilon](0\,|\,\Omega)\neq0.
\non
\ena
Taking  this quantity as a normalization constant we define the sigma function as follows.

\begin{defn} The sigma function associated with  $(X, \{\alpha_i,\beta_i\}, p_\infty, z, e,\{du_{w_i}\}, \widehat{\omega})$ is defined by 
\bea
&&
\sigma[\varepsilon](u)=\exp(\frac{1}{2}{}^tu\eta_1\omega_1^{-1} u)
\frac{\theta[\varepsilon]((2\omega_1)^{-1}u\,|\,\Omega)}{c_0\partial_{A_0}\theta[\varepsilon](0\,|\,\Omega)},
\label{def-sigma}
\ena
where $c_0=\pm1$ is given by (\ref{signature}).
\end{defn}

%%%%%%%%%%%%%%%%%%%%%%%%%%%%%%%%%%%%%%%%%%%%%%%%%%%%%%%%%%%%%%%%%%%%%%%%%%%%%%%%%%%%%%
\subsection{Modular invariance}

\begin{theorem}
The function $\sigma[\varepsilon](u)$ is independent of the choice of $\{\alpha_i,\beta_i\}$ and has the series 
expansion of the form 
\bea
&&
\sigma[\varepsilon](u)= s_\lambda(t)|_{t_{w_i}=u_{w_i}}+\text{higher weight terms},
\label{sigma-expansion}
\ena
where $\lambda$ is given by (\ref{lambda}).
\end{theorem} 
\vskip2mm
\noindent
{\it Proof.} The series expansion (\ref{sigma-expansion}) follows from Theorem \ref{expansion-theta}.
The invariance under the change of the canonical homology basis is proved based on the 
following transformation formula for the theta function in \cite{I}.

Let 
\bea
&&
M=\left(\begin{array}{cc}A&B\\C&D\end{array}\right)
\non
\ena
be an element of the symplectic group $Sp(2g,{\mathbb Z})$, where $A,B,C,D$ are integral 
matrices of degree $g$, and $z={}^t(z_1,...,z_g)$.
We set
\bea
&&
\tilde{\Omega}=(A\Omega+B)(C\Omega+D)^{-1},\qquad
\tilde{z}={}^t(C\Omega+D)^{-1}z,
\label{tilde-Omega}
\\
&&
\tilde{\varepsilon}=\left(\begin{array}{cc}D&-C\\-B&A\end{array}\right)\varepsilon+
\frac{1}{2}\text{diag}\left(\begin{array}{c}C{}^tD\\A{}^t B\end{array}\right),
\label{tilde-epsilon}
\ena
where $\text{diag}(\cdot)$ denotes the column vector whose components are the diagonal 
entries of the matrices in $(\cdot)$. Then 
\bea
&&
\theta[\tilde{\varepsilon}](\tilde{z}\,|\,\tilde{\Omega})=\gamma (\det(C\Omega+D))^{\frac{1}{2}}
e^{\pi i {}^tz (C\Omega+D)^{-1} C z}\theta[\varepsilon](\,z\,|\,\Omega),
\label{theta-mod-trf}
\ena
where $\gamma$ is some $8$-th root of unity.

Let $\alpha={}^t(\alpha_1,...,\alpha_g)$, $\beta={}^t(\beta_1,...,\beta_g)$. 
Let 
\bea
&&
T=\left(\begin{array}{cc}D&C\\B&A\end{array}\right)
\non
\ena
be an element of  $Sp(2g,{\mathbb Z})$.
We change the canonical homology basis by $T$:
\bea
&&
\left(\begin{array}{c}\tilde{\alpha} \\ \tilde{\beta}\end{array}\right)=T
\left(\begin{array}{c}\alpha \\ \beta\end{array}\right).
\label{change-basis}
\ena

Notice that $T$ is an element of  $Sp(2g,{\mathbb Z})$ if and only if $M$ is an element 
of  $Sp(2g,{\mathbb Z})$. Let $\tilde{\omega}_i, \tilde{\eta}_i$ be the period matrices 
corresponding to $\{\tilde{\alpha}_i,\tilde{\beta}_i\}$. Since the bilinear differential 
$\widehat{\omega}(p_1,p_2)$ does not 
depend on the choice of a canonical homology basis we have, by computation, 
\bea
&&
\tilde{\omega}_1=\omega_1{}^tD+\omega_2{}^tC,\qquad
\tilde{\omega}_2=\omega_1{}^tB+\omega_2{}^tA,
\non
\\
&&
\,
\tilde{\eta}_1=\eta_1{}^tD+\eta_2{}^tC,\qquad\,\,
\tilde{\eta}_2=\eta_1{}^tB+\eta_2{}^tA.
\non
\ena
Then the normalized period matrix $\tilde{\Omega}$ corresponding to  $\{\tilde{\alpha}_i,\tilde{\beta}_i\}$   becomes
\bea
&&
\tilde{\Omega}={}^t\tilde{\Omega}={}^t(\tilde{\omega}_1^{-1}\tilde{\omega}_2)=
(A\Omega+B)(C\Omega+D)^{-1}.
\non
\ena
Using Riemann's vanishing theorem we see that Riemann's constant $\delta$ changes to (see also \cite{F})
\bea
&&
\tilde{\delta}={}^t(C\Omega+D)^{-1}\delta-\tilde{\Omega}\zeta'-\zeta'',
\non
\ena
where $\zeta',\zeta''\in (1/2{\mathbb Z}^g)/{\mathbb Z}^g$ are given by 
\bea
&&
\left(\begin{array}{c}\zeta' \\ \zeta''\end{array}\right)=
\frac{1}{2}\text{diag}\left(\begin{array}{c}C{}^tD\\A{}^t B\end{array}\right).
\non
\ena
Then $e$ transforms to $\tilde{e}=\Omega\tilde{\varepsilon}'+\tilde{\varepsilon}''$ with 
$\varepsilon$ being given by (\ref{tilde-epsilon}).

Notice that 
\bea
&&
(2\tilde{\omega}_1)^{-1}={}^t(C\Omega+D)^{-1}(2\omega_1)^{-1}.
\non
\ena
Therefore if we change the canonical homology basis to $\{\tilde{\alpha}_i,\tilde{\beta}_i\}$ 
by (\ref{change-basis}), the theta function $\theta[\varepsilon]((2\omega_1)^{-1}u\,|\Omega)$
changes to 
\bea
&&
\theta[\tilde{\varepsilon}]({}^t(C\Omega+D)^{-1}(2\omega_1)^{-1} u\,|\tilde{\Omega}),
\non
\ena
where $\tilde{\Omega}$, $\tilde{\varepsilon}$ are given by (\ref{tilde-Omega}), (\ref{tilde-epsilon}).
Thus the formula (\ref{theta-mod-trf}) can be applied. Then we have

\bea
\theta[\tilde{\varepsilon}]((2{\tilde{\omega}}_1)^{-1} u\,|\tilde{\Omega})&=&
\gamma \left(\det(C\Omega+D)\right)^{1/2}
e^{\pi i{}^t u {}^t(2\omega_1)^{-1}(C\Omega+D)^{-1} C(2\omega_1)^{-1} u}
\non
\\
&&
\times\theta[\varepsilon]((2\omega_1)^{-1} u\,|\Omega).
\label{trf-modular1}
\ena
Applying $\partial_{A_0}$ to (\ref{trf-modular1}) and 
set $u=0$. Then, using (1) or (2) of Corollary \ref{refined-RST}, we have 
\bea
&&
\partial_{A_0}
\theta[\tilde{\varepsilon}](0\,|\tilde{\Omega})
=\gamma  \left(\det(C\Omega+D)\right)^{1/2}
\partial_{A_0}
\theta[\varepsilon](0\,|\Omega).
\label{trf-modular2}
\ena
On the other hand, as shown in \cite{BEL1}, we have
\bea
&&
\frac{1}{2}{}^t u{\tilde{\eta}}_1{\tilde{\omega}}_1^{-1} u
=\frac{1}{2}{}^t u \eta_1 \omega_1^{-1} u
-\pi {}^t u {}^t(2\omega_1)^{-1}(C\Omega+D)^{-1}C(2\omega_1)^{-1} u.
\label{trf-modular3}
\ena
By (\ref{trf-modular1}),  (\ref{trf-modular2}),  (\ref{trf-modular3}) we have
\bea
&&
\exp(\frac{1}{2}{}^tu{\tilde{\eta}}_1{\tilde{\omega}}_1^{-1} u)
\frac{\theta[\tilde{\varepsilon}]((2{\tilde{\omega}}_1)^{-1}u\,|\,\tilde{\Omega})}{\partial_{A_0}\theta[\tilde{\varepsilon}](0\,|\,\tilde{\Omega})}
\non
\\
&&
\hskip50mm
=\exp(\frac{1}{2}{}^tu\eta_1\omega_1^{-1} u)
\frac{\theta[\varepsilon]((2\omega_1)^{-1}u\,|\,\Omega)}{\partial_{A_0}\theta[\varepsilon](0\,|\,\Omega)},
\non
\ena
which shows that $\sigma[\varepsilon](u)$ does not depend on the choice of 
canonical homology basis. $\Box$

\vskip10mm
\noindent
{\large {\bf Acknowledgments}} 
\vskip3mm
\noindent
This research is supported by JSPS Grant-in-Aid for Scientific Research (C) 
No.23540245. A part of results of this paper was reported at the conference 
"Algebraic topology and abelian functions" held at Steklov Mathematical Institute, Moscow, in June 2013,  
at the conference "Geometry day" held at Sobolev Institute of Mathematics, Novosibirsk in August 2013 and 
at the conference "Algebraic curves with symmetries, their Jacobians and integrable dynamical 
systems" at National University of Kyiv-Mohyla Academy, Kiev in July 2014. I would like to thank 
organizers for invitations and kind hospitality at the conferences. I would also like to thank Harry Braden, 
Victor Buchstaber, Victor  Enolski,  Dmitry Korotkin,  Dmitry Leykin, Andrey Mironov and Alexander Veselov for their interests  in this work.

\end{document}